\documentclass{svjour3}                     
\usepackage{amssymb}
\usepackage{graphicx,epsfig}
\usepackage{natbib}
\usepackage{mathptmx}      
\bibpunct{(}{)}{;}{a}{}{,}
\makeatletter
\renewcommand*{\@biblabel}[1]{\hfill}
\makeatother
\newcommand{\ion}[2]{#1\,{\sc{#2}}}
\def\lsim{\mathrel{\rlap{\lower4pt\hbox{\hskip1pt$\sim$}}
    \raise1pt\hbox{$<$}}}                
\def\gsim{\mathrel{\rlap{\lower4pt\hbox{\hskip1pt$\sim$}}
    \raise1pt\hbox{$>$}}}                

\journalname{SSRv}

\begin{document}

\title{FUV and X-ray absorption in the Warm-Hot Intergalactic Medium }

\author{P. Richter \and
        F.B.S. Paerels \and
        J.S. Kaastra}

\authorrunning{Richter et al.}
\titlerunning{FUV and X-ray absorption of the WHIM}

\institute{P. Richter \at Institut f\"ur Physik, Universit\"at Potsdam,
                  Am Neuen Palais 10, D-14469 Potsdam, Germany\\
                  \email{prichter@astro.physik.uni-potsdam.de}
           \and F.B.S. Paerels \at Department of Astronomy and Columbia Astrophysics Laboratory, Columbia University, 550 West 120th Street, New York, NY 10027, USA
           \and J.S. Kaastra \at  SRON, Sorbonnelaan 2, 3584 CA Utrecht, The Netherlands
	     }

\date{Received: 20 September 2007; Accepted: 21 September 2007}

\maketitle

\begin{abstract}
The Warm-Hot Intergalactic Medium (WHIM) arises from shock-heated 
gas collapsing in large-scale filaments and probably harbours
a substantial fraction of the baryons in the local Universe.
Absorption-line measurements in the ultraviolet (UV) and 
in the X-ray band currently represent the best method to study 
the WHIM at low redshifts. 
We here describe the physical properties of the WHIM and the concepts
behind WHIM absorption line measurements of \ion{H}{i} and 
high ions such as \ion{O}{vi}, \ion{O}{vii}, and \ion{O}{viii} in
the far-ultraviolet and X-ray band. We review results of recent WHIM absorption 
line studies carried out with UV and X-ray satellites such as FUSE, 
HST, Chandra, and XMM-Newton and discuss their implications for
our knowledge of the WHIM.

\end{abstract}

\section{Introduction}
\label{Introduction} 

As recent cosmological simulations imply, the 
temperature of the intergalactic medium (IGM)
undergoes a significant change from
high to low redshifts parallel to the proceeding of
large-scale structure formation in the Universe
(e.g., \citealt{cen1999}; \citealt{dave2001}).
As a result, a substantial fraction of the baryonic 
matter in the local Universe is expected to reside in 
the so-called Warm-Hot Intergalactic Medium (WHIM).
The WHIM represents a low-density ($n_{\rm H}\sim10^{-6}-10^{-4}$ cm$^{-3}$),
high-temperature ($T\sim 10^5-10^7$ K) plasma that primarily is made of
protons, electrons, \ion{He}{ii}, and \ion{He}{iii}, together with traces of some
highly-ionised heavy elements. The WHIM is believed to 
emerge from intergalactic gas that is shock-heated to 
high temperatures as the medium is collapsing under the 
action of gravity in large-scale filaments (e.g., \citealt{valageas2002}).
In this scenario, part of the warm (photoionised) intergalactic medium that
gives rise to the Ly\,$\alpha$ forest in the spectra of distant 
quasars (QSO) is falling in to the potential wells of the 
increasingly pronounced filaments, gains energy
(through gravity), and is heated to high temperatures by shocks that
run through the plasma.

Because of the low density and the high degree of ionisation, direct 
observations of the shock-heated and collisionally ionised 
WHIM are challenging with current instrumentation 
(in contrast to the photoionised IGM, which is easily observable 
through the Ly\,$\alpha$ forest).
Diffuse emission from the WHIM plasma must have a very low surface brightness
and its detection awaits UV and X-ray observatories more sensitive 
than currently available (see, e.g., \citealt{fang2005}; \citealt{kawahara2006}).
The most promising approach to study the WHIM with observations at low
redshift is to 
search for absorption features from the WHIM in FUV and in the X-ray regime
in the spectra of quasars, active galactic nuclei (AGN) and other 
suited extragalactic background sources.
As the WHIM represents a highly-ionised plasma, the 
most important WHIM absorption lines are those originating from
the electronic transitions of 
high-ionisation state ions
(hereafter referred to as "high ions") of abundant heavy elements 
such as oxygen and carbon. 
Among these, five-times ionised oxygen (\ion{O}{vi}) is 
the most valuable high ion to trace the WHIM at temperatures of
$T\sim 3\times 10^5$ K in the FUV regime. In the X-ray band, the 
\ion{O}{vii} and \ion{O}{viii} transitions represent the
key observables to trace the WHIM at higher temperatures in the
range $3\times 10^5 < T < 10^7$.
In addition to the spectral signatures of high ions of heavy 
elements the search for broad and shallow Ly\,$\alpha$ absorption from the 
tiny fraction of neutral hydrogen in the WHIM represents another
possibility to identify and study the most massive WHIM filaments 
in the intergalactic medium with FUV absorption spectroscopy.
Finally, for the interpretation of the observed WHIM absorption features 
in UV and X-ray spectra the comparison between real data and 
artificial spectra generated by numerical simulations that include
realistic gas physics is of great importance to identify 
possible pitfalls related to technical and physical 
issues such as limited signal-to-noise ratios and spectral resolution, 
line-broadening mechanisms, non-equilibrium conditions, and others.

In this chapter, we review the physics and methodology of the UV 
and X-ray absorption measurements of warm-hot intergalactic gas 
at low redshift and summarise the results of recent 
observations obtained with space-based observatories. The
outline of this chapter is the following.
The ionisation conditions of the WHIM and the most important
absorption signatures of this gas in the UV and X-ray band 
are presented in Sect.\,2. Recent UV absorption measurements 
of the WHIM at low redshift are discussed in Sect.\,3.
Similarly, measurements of the WHIM in the X-ray are presented
in Sect.\,4. In Sect.\,5 we compare the results from WHIM
observations with predictions from numerical simulations and
give an overview of WHIM measurements at high redshift. 
Finally, some concluding remarks are given in Sect.\,6.


\section{Physical properties of the WHIM}
\label{Physical properties of the WHIM}

\subsection{WHIM ionisation conditions} 
\label{WHIM ionisation conditions}

The occurrence and characteristics of the WHIM absorption 
signatures in the FUV and X-ray band are determined to a
high degree by the ionisation conditions in the gas.
We briefly discuss the WHIM ionisation properties,
as this is crucial for interpretation of the WHIM absorption
lines in FUV and X-ray spectra that arise in such warm-hot
gas.
Generally, there are two processes that determine the 
ionisation state of warm-hot gas in the intergalactic medium: 
collisional ionisation caused by the high temperature of the
gas in collapsed structures and 
photoionisation by the cosmic FUV background. 

\subsubsection{Hydrogen}
\label{Hydrogen}

By far most of the mass of the WHIM is in the form of ionised
hydrogen. Therefore, understanding the processes that lead
to the ionisation of hydrogen is essential for the interpretation
of WHIM absorption lines and for a reliable estimate of the 
baryon content of warm-hot intergalactic gas.
The ionisation potential of neutral hydrogen is $13.6$ eV and thus
both ionisation by particle collisions and ionisation by high-energy
photons contribute to the ionisation of \ion{H}{i} in warm-hot gas.
We start with collisional ionisation, which is believed to
dominate the ionisation of hydrogen at temperatures $>10^5$ K.

In collisional ionisation equilibrium (CIE) -- the most
simple approach to characterise the ionisation conditions 
in low-density, high-temperature plasmas -- the ionisation fraction
depends only on the gas temperature. If we ignore any charge-exchange
reactions (which is justified in case of hydrogen), the neutral hydrogen
fraction in CIE is simply the ratio between the recombination coefficient
$\alpha_{\rm H}(T)$ and the collisional ionisation coefficient
$\beta_{\rm H}(T)$:
\begin{equation}
f_{\mbox{H\,{\scriptsize I}}, \rm coll}=\frac{\alpha_{\rm H}(T)}{\beta_{\rm H}(T)}.
\end{equation}

Above gas temperatures of $\sim 1.5 \times 10^4$ K collisions
by thermal electrons efficiently ionise hydrogen to a high degree,
and already at $T\sim 3 \times 10^4$ K the neutral hydrogen fraction
in the gas is less then one percent. For the temperature range 
that is characteristic for the WHIM, $T=10^5-10^7$ K, 
one can approximate the ionisation fraction
in a collisional ionisation equilibrium in the way
\begin{equation}
\log\,f_{\mbox{H\,{\scriptsize I}}, \rm coll}
\approx 13.9 - 5.4\,{\rm log}\,T + 0.33\,
({\rm log}\,T)^2.
\end{equation}
where $T$ is in units K (\citealt{richter2006a}; \citealt{sutherland1993}).
Thus, for WHIM gas with 
$T=10^6$ K the neutral hydrogen
fraction in the gas in CIE is only $\sim 2.4 \times 10^{-7}$.

\begin{figure}    
\begin{center}
\includegraphics[width=\hsize]{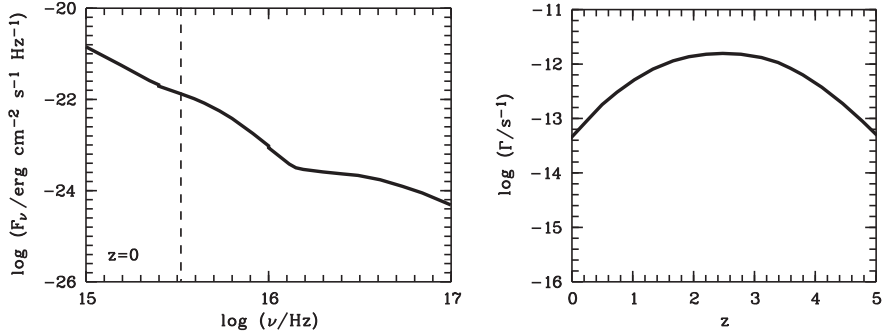}
\caption{
{\it Left panel:} Spectral shape of the metagalactic UV background at $z=0$
(from \protect\citealt{haardt1996}). 
Plotted is the flux of photons ($F_{\nu}=4 \pi J_{\nu}$) against 
the frequency $\nu$. 
The hydrogen ionisation edge is indicated with a dashed line.
{\it Right panel:} Redshift-dependence of the hydrogen photoionisation rate
$\Gamma$ from the UV background for the range $z=0$ to $z=5$.
Adapted from \protect\citet{haardt1996}.
}
\label{fig:fig1}
\end{center}
\end{figure}

Next to particle collisions,
photons with energies $>13.6$ eV contribute to the ionisation 
of the WHIM, in particular in the low-temperature WHIM tail at 
$\sim 10^5$ K and below. Such ionising photons in intergalactic
space are indeed provided by the metagalactic ultraviolet (UV) background,
originating from the hard radiation emitted by QSOs and AGN.
Fig.~\ref{fig:fig1} shows the spectral shape of the UV background at $z=0$ (left
panel) and the redshift-dependence of the hydrogen photoionisation rate 
from the UV background (right panel) based on the models by 
\citet{haardt1996}.

\begin{figure}    
\begin{center}
\includegraphics[width=0.67\hsize]{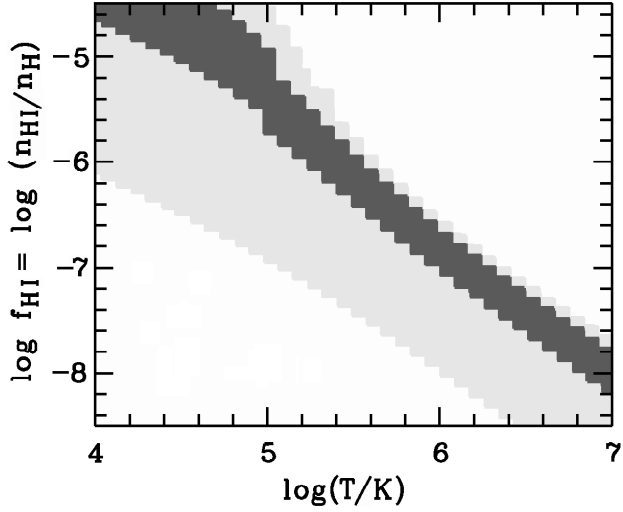}
\caption{
The neutral hydrogen fraction,
log $f_{\rm H\,I}
={\rm log} (n_{\rm H\,I}/n_{\rm H})$,
in a WHIM simulation (photoionisation$+$collisional ionisation),
is plotted as a function of the gas temperature, log $T$.
The light gray shaded indicates cells in the density range
log $n_{\rm H}=-5$ to $-7$. The dark gray shaded area refers to
cells that have log $n_{\rm H}=-5.3$ to $-5.6$, thus a density
range that is characteristic for WHIM absorbers. 
Adapted from \protect\citet{richter2006b}.
}
\label{fig:fig2}
\end{center}
\end{figure}

Considering photoionisation, one generally can write for 
the neutral-hydrogen fraction in the gas:
\begin{equation}
f_{\mbox{H\,{\scriptsize I}},\rm photo} = 
\frac{n_{\mathrm e}\,\alpha_{\rm H}(T)}{\Gamma_{\mbox{H\,{\scriptsize I}}}},
\end{equation}
where $\alpha_{\rm H}(T)$ denotes the temperature-dependent 
recombination rate of hydrogen, $n_{\mathrm e}$ is the electron density, 
and $\Gamma_{\mbox{H\,{\scriptsize I}}}$
is the photoionisation rate. $\Gamma_{\mbox{H\,{\scriptsize I}}}$ 
depends on the ambient ionising radiation field $J_{\nu}$ (in units 
erg\,cm$^{-2}$\,s$^{-1}$\,Hz$^{-1}$ sr$^{-1}$) in the WHIM 
provided by the metagalactic UV background (see Fig.~\ref{fig:fig1}):
\begin{equation}
\Gamma_{\mbox{H\,{\scriptsize I}}} = 4\pi\,\int\limits_{\displaystyle{\nu_{\rm L}}}^{\displaystyle{\infty}}
\frac{\sigma_{\nu} J_{\nu}}{{\mathrm h}\nu}\,{\mathrm d}\nu
\approx 2.5\times10^{-14}\,J_{-23}\,{\rm s}^{-1}.
\end{equation}
Here, $\nu_{\rm L}$ is the frequency at the Lyman limit
and ${\sigma_{\nu}}$ denotes the photoionisation cross section
of hydrogen, which scales with $\nu^{-3}$ for frequencies
larger that $\nu_{\rm L}$ (see \citealt{kaastra2008} - Chapter 9, this volume).
We have introduced the dimensionless scaling factor 
$J_{-23}$ which gives the metagalactic UV radiation 
intensity at the Lyman limit in units 
$10^{-23}$ erg\,cm$^{-2}$\,s$^{-1}$\,Hz$^{-1}$\,
sr$^{-1}$. For $z=0$ we have $J_{-23}\sim 1-2$, while for
$z=3$ the value for $J_{-23}$ is $\sim 80$, thus significantly
higher \citep{haardt1996}. 
Assuming $n_{\rm e} = n_{\rm H}$ and inserting a proper
function for $\alpha_{\rm H}(T)$, we finally can write for the
logarithmic neutral hydrogen fraction in a purely 
photoionised WHIM plasma 
\begin{equation}
{\rm log}\,f_{\mbox{H\,{\scriptsize I}}, \rm photo}
\approx {\rm log}\,\left( \frac{16\,n_{\rm H}
\,T_4^{-0.76}}{J_{-23}}\right),
\end{equation}
where $n_{\rm H}$ is the hydrogen volume density 
in units cm$^{-3}$ and $T_4$ is the temperature in
units $10^4$ K. Thus, for purely photoionised intergalactic
gas at $z=0$ with $n=5\times 10^{-6}$ and $T=10^6$ K
we find that the neutral hydrogen fraction
is $ f_{\mbox{H\,{\scriptsize I}}, \rm photo} \sim 2.4 \times 10^{-6}$.
This is ten times higher than for CIE,
indicating that collisions dominate the ionisation fraction
of hydrogen in intermediate and high-temperature WHIM regions.
However, note that at lower temperatures near $T=10^5$ K 
at the same density we have $f_{\mbox{H\,{\scriptsize I}}, \rm photo} 
\sim f_{\mbox{H\,{\scriptsize I}}, \rm coll}$. Since this is the WHIM temperature
regime preferentially detected by UV absorption features
(e.g., \ion{O}{vi} and broad Ly\,$\alpha$), photoionisation
is important and needs to be accounted for when it comes to
the interpretation of WHIM absorbers observed in the FUV.
From a WHIM simulation at $z=0$ 
including both collisional ionisation and photoionisation
(\citealt{richter2006b}; see Fig.~\ref{fig:fig2}) 
find the following empirical relation between neutral hydrogen 
 fraction and gas temperature for a WHIM
density range between log $n_{\rm H}=-5.3$ and $-5.6$:
\begin{equation}
{\rm log}\,f_{\mbox{H\,{\scriptsize I}}}
\approx 0.75 - 1.25\,{\rm log}\,T.
\end{equation}

This equation may serve as a thumb rule to estimate ionisation
fractions in WHIM absorbers at $z=0$ if the gas temperature is
known (e.g., from measurements of the line widths; see Sect.\,2.2.1).

\begin{figure}    
\begin{center}
\includegraphics[width=\hsize]{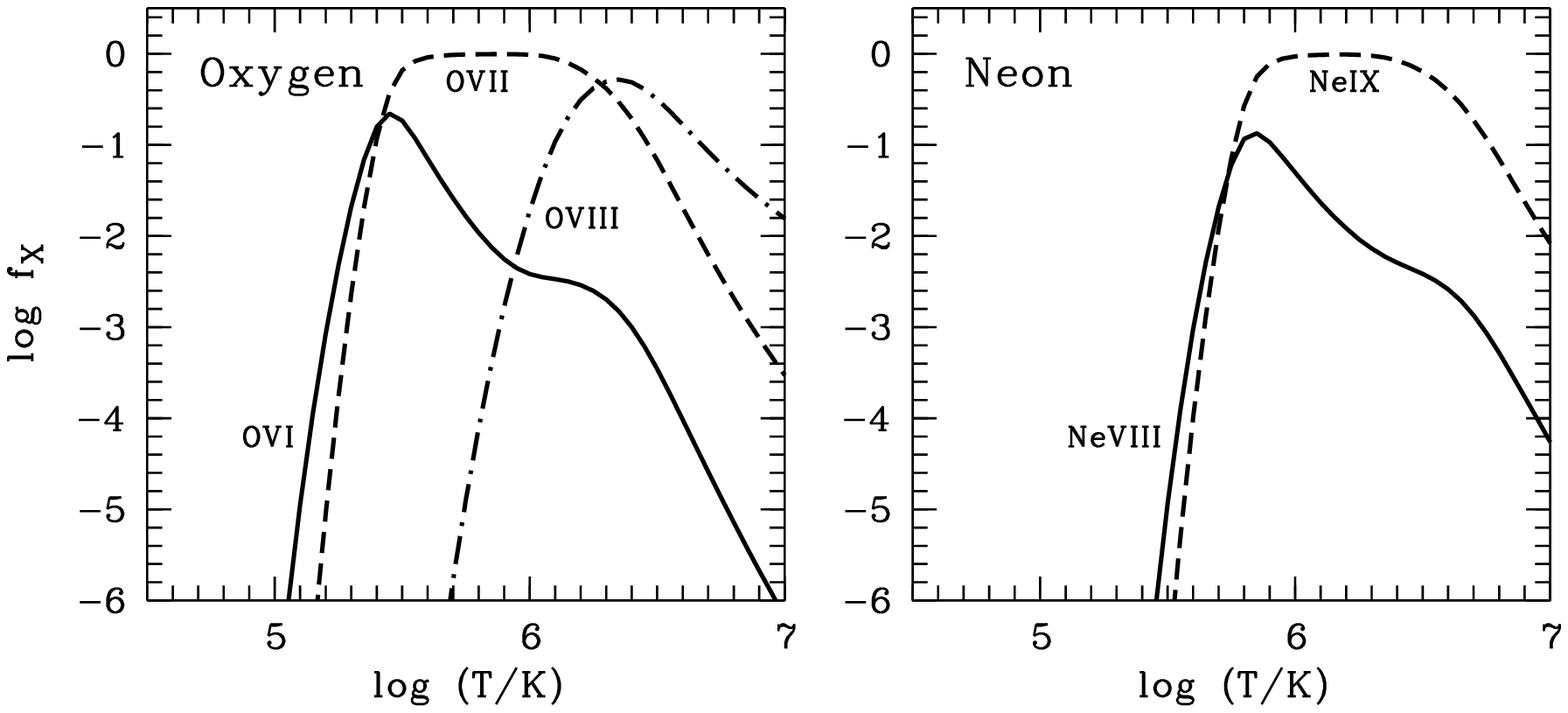}
\caption{
CIE ion fractions 
of selected high ions of oxygen (\ion{O}{vi}, \ion{O}{vii}, \ion{O}{viii}; left panel)
and neon (\ion{Ne}{viii}, \ion{Ne}{ix}; right panel) in the WHIM temperature range 
log ($T$/K$)=4.5-7.0$, based on calculations
by \citet{sutherland1993}.
}
\label{fig:fig3}
\end{center}
\end{figure}

\subsubsection{Oxygen and other metals}
\label{Oxygen and other metals}

While hydrogen provides most of the mass in the WHIM, the
most important diagnostic lines to study this gas phase
are from highly ionised metals such as oxygen, 
neon, carbon, magnesium, and others. Therefore, the understanding of the
ionisation properties of the observed high ions of these 
elements is as important as for hydrogen. As for hydrogen,
both collisional ionisation and photoionisation need to
be considered.
With its single electron, hydrogen can only be either neutral
or fully ionised. 
Heavy elements, in contrast, have several electrons available and
are -- even at very high temperatures -- usually only partly
ionised. Thus, electronic transitions exist for such
highly-ionised metals ("high ions") in warm-hot gas.
Of particular importance for observations of the WHIM are the high
ionisation states of oxygen, \ion{O}{vi}, \ion{O}{vii},
and \ion{O}{viii}, as they have strong electronic transitions
in the UV (\ion{O}{vi}) and at X-ray wavelengths (\ion{O}{vii} 
\& \ion{O}{viii}) and oxygen is a relatively abundant element.
Another important metal for observing
warm-hot gas in the UV and X-ray band is neon (\ion{Ne}{vii},
\ion{Ne}{viii}, \ion{Ne}{ix}, \ion{Ne}{x}). In collisional ionisation equilibrium,
the ionisation state of these elements is determined solely by the
temperature of the gas. For each element, the ionisation fractions
of the ionisation states (e.g., four-times vs. five-times
ionised) then are characterised by the respective ionisation potentials (IPs)
of the individual ionisation levels.
For instance, at $T\sim 1-3 \times 10^5$ K, a significant fraction of the oxygen 
is five-times ionised (O$^{+5}$ or \ion{O}{vi}, IP$=138$ eV).
Six-times ionised oxygen (O$^{+6}$ or \ion{O}{vii}, IP$=739$ eV) and seven-times 
ionised oxygen (O$^{+7}$ or \ion{O}{viii}, IP$=871$ eV) predominantly exist at 
higher temperatures in the range $3\times 10^5 - 3\times 10^6$ K
and $3\times 10^6 - 10^7$ K, respectively.
Fig.~\ref{fig:fig3} shows the ionisation fractions of the most important high 
ions of oxygen and neon, based on the CIE calculations of \citet{sutherland1993};
see also \citealt{kaastra2008} - Chapter 9, this volume.
High ions of other elements such as carbon, nitrogen, silicon and magnesium are 
less important for WHIM observations as their observable transitions 
trace lower temperature gas (e.g., \ion{C}{iv}, \ion{Si}{iv}) or the 
abundance of these elements in the intergalactic medium are too low.
It is important to note at this point, that the discussed relation between 
ionisation state/fraction and gas temperature explicitly assumes that the
gas is in an ionisation {\it equilibrium}. This may not be
generally the case in the WHIM, however, as the densities 
are generally very low. For instance, under particular
non-equilibrium conditions the timescales for cooling, recombination,
and ion/electron equilibration
may differ significantly from each other (see for instance \citealt{bykov2008} - Chapter 8, this volume). In such a case, 
the presence of high ions such as \ion{O}{vi} and/or measured high-ion
ratios would {\it not} serve as a reliable 
"thermometer" for the WHIM gas. In addition, WHIM filaments
most likely neither are isothermal nor do they have a constant
particle density. In fact, as WHIM simulations demonstrate,
WHIM absorbers seem to represent a mix of cooler 
photoionised and hotter collisionally ionised gas
with a substantial intrinsic density range. The absorption
features from high ions arising in such a multi-phase 
medium therefore are generally difficult to interpret
in terms of physical conditions and baryon budget.

In view of the high energies required to produce the high ions of oxygen
and neon in combination with the spectral shape of the 
metagalactic background radiation (see Fig.~\ref{fig:fig1}), photoionisation of 
high metal ions in the WHIM is less important than for hydrogen.
However, for \ion{O}{vi} photoionisation is important at low redshifts
in WHIM regions with very low densities or in systems located
close to a strong local radiation source (e.g., in \ion{O}{vi} systems 
associated with the background QSO). Note that at high redshift,
most of the intervening \ion{O}{vi} appears to be photoionised  
owing to the significantly higher intensity of the 
metagalactic background radiation
in the early Universe (see Sect.\,5.2).

\begin{table}
\caption[]
{Data on O and Ne high ions having observable absorption lines}
\begin{tabular}{lrrrlr}
\hline\hline
Ion  & [X/H]$^1$ & Ionisation     & Absorption  & Band & CIE temperature$^2$ \\
     &           & potential [eV] & lines [\AA] &      & range [$10^6$ K] \\
\hline
\ion{O}{vi} & $-3.34$  & $138$ & $1031.926$ & FUV    & $0.2-0.5$ \\
            &          &          & $1037.617$ &        &           \\
\ion{O}{vii} & $-3.34$  & $739$ & $21.602$   & X-ray  & $0.3-3.0$   \\
\ion{O}{viii} & $-3.34$  & $871$ & $18.969$   & X-ray  & $1.0-10.0$    \\
\hline
\ion{Ne}{viii} & $-4.16$  & $239$ & $770.409$ & EUV    & $0.5-1.3$ \\
          &          &             & $780.324$ &        &           \\
\ion{Ne}{ix} & $-4.16$  & $1196$ & $13.447$  & X-ray  & $0.6-6.3$   \\
\hline
\end{tabular}
\noindent\\
$^1$ [X/H] is the $\log$ of the number density of element X relative to hydrogen for Solar abundances, taken here from \citet{asplund2004}.\\
$^2$ CIE models from \citet{sutherland1993}.
\end{table}

\subsection{WHIM absorption signatures in the UV and X-ray band}
\label{WHIM absorption signatures in the UV and X-ray band}

\subsubsection{UV absorption}
\label{UV absorption}

As indicated in the previous subsection, five-times ionised oxygen (\ion{O}{vi}) is
by far the most important high ion to trace the WHIM at temperatures of
$T\sim 3\times 10^5$ K in the ultraviolet regime (assuming CIE, see above). 
Oxygen is a relatively abundant element and the 
two lithium-like $(1s^22s)\,^2S_{1/2}\rightarrow(1s^22p)\,^2P_{1/2,3/2}$
electronic transitions of \ion{O}{vi} located in the FUV at $1031.9$ and $1037.6$ \AA\, have
large oscillator strengths ($f_{1031}=0.133, f_{1037}=0.066$).
Next to \ion{O}{vi},
\ion{Ne}{viii} traces WHIM gas near $T\sim 7\times10^5$ K (in collisional
ionisation equilibrium) and thus
is possibly suited to complement the \ion{O}{vi} measurements
of the WHIM in a higher temperature regime.
The two available \ion{Ne}{viii} lines are located in the extreme
ultraviolet (EUV) at $770.4$ \AA\, ($f_{770}=0.103$) and
$780.3$ \AA\, ($f_{780}=0.051$), allowing us to trace 
high-column density WHIM absorbers at redshifts $z>0.18$ 
with current FUV satellites such as FUSE.
However, as the cosmic abundance of \ion{Ne}{viii} is relatively low,
\ion{Ne}{viii} is not expected to be a particularly sensitive tracer of the
WHIM at the S/N levels achievable with current UV
spectrographs. The same argument holds for the high ion \ion{Mg}{x}, 
which has two transitions in the EUV at even lower wavelengths 
($\lambda\lambda$ 609.8, 624.9~\AA). So far, only \ion{O}{vi}
and in one case \ion{Ne}{viii} has been observed in the WHIM
at low redshift (see Sect.\,3.2). Note that WHIM absorption features by
\ion{O}{vi} (and \ion{Ne}{viii}) are mostly unsaturated 
and the line profiles are fully or nearly resolved by current 
UV instruments such as FUSE and STIS, which provide spectral
resolutions of $R=\lambda/\Delta \lambda \approx 20,000$ and $45,000$, 
respectively. 
Table 1 summarises physical parameters
of O and Ne high ions and their observable transitions in the UV and X-ray bands.

Four-times ionised nitrogen (\ion{N}{v}; I.P. is $98$ eV) also
is believed to trace predominantly collisionally ionised gas at temperatures
near $T\sim 2\times10^5$ K, but its lower cosmic abundance together with its
deficiency in low metallicity environments due to nucleosynthesis effects
(e.g., \citealt{pettini2002}) makes it very difficult to detect in the WHIM.
Other available strong high-ion transitions in the UV
from carbon (\ion{C}{iv} $\lambda\lambda$ 1548.2, 1550.8~\AA) and
silicon (\ion{Si}{iv} $\lambda\lambda$ 1393.8, 1402.8~\AA) are believed
to trace mainly photoionised gas at temperatures $T<10^5$ K,
but not the shock-heated warm-hot gas at higher temperatures.

Next to high-ion absorption from heavy elements,
recent UV observations (\citealt{richter2004}; \citealt{sembach2004}; \citealt{lehner2007}) have indicated that WHIM filaments 
can be detected in Ly\,$\alpha$ absorption of neutral hydrogen.
Although the vast majority of the hydrogen in the WHIM
is ionised (by collisional processes and UV radiation), a
tiny fraction ($f_{\rm H\,I}<10^{-5}$, typically) of neutral
hydrogen is expected to be present. Depending on the total
gas column density of a WHIM absorber and its
temperature, weak \ion{H}{i} Ly\,$\alpha$ absorption
at column densities $12.5\leq$ log $N$(\ion{H}{i})$\leq 14.0$
may arise from WHIM filaments and could be used to
trace the ionised hydrogen component.
The Ly\,$\alpha$ absorption from WHIM filaments is
very broad due to thermal line
broadening, resulting in large Doppler parameters
of $b>40$ km\,s$^{-1}$. Such lines are generally difficult
to detect, as they are broad and shallow. High resolution,
high S/N FUV spectra of QSOs with smooth background continua
are required to successfully search for broad Ly\,$\alpha$
absorption in the low-redshift WHIM. STIS installed on the HST
is the only instrument that has provided such data,
but due to the instrumental limitations of space-based
observatories, the number of QSO spectra adequate for searching
for WHIM broad Ly\,$\alpha$ absorption
(in the following abbreviated as "BLA") is very limited. 

The $b$ values of the BLAs are assumed to be composed
of a thermal component, $b_{\rm th}$, and a non-thermal
component, $b_{\rm nt}$, in the way that
\begin{equation}
b=\sqrt{b_{\rm th}\,^2+b_{\rm nt}\,^2}.
\end{equation}

The non-thermal component may include processes like
macroscopic turbulence, unresolved velocity-components,
and others (see \citealt{richter2006a} for a detailed discussion).
The contribution of the thermal component to $b$
depends on the gas temperature:
\begin{equation}
b_{\rm th} = \sqrt \frac{2kT}{m} \approx 0.13 \, \sqrt
\frac{T}{A}\, \rm{km\,s}^{-1},
\end{equation}

where $T$ is in K, $k$ is the Boltzmann constant, $m$ is the particle mass,
and $A$ is the atomic weight.
For the shock-heated WHIM gas with log $T\geq5$ one thus expects $b_{\rm th}\geq40$
km\,s$^{-1}$. The non-thermal broadening mechanisms
are expected to contribute to some degree to the total
$b$ values in WHIM absorbers (see \citealt{richter2006a}), so that
the measured $b$ value of a BLA provides only an upper limit for the
temperature of the gas.

\subsubsection{X-ray absorption}
\label{X-ray absorption}

The highest ionisation phase of the WHIM will produce and absorb line 
radiation primarily in the He- and H-like ions of the low-$Z$ elements 
(C, N, O, Ne), and possibly in the L-shell ions of Fe. In practice, much 
of the attention is focused on oxygen, because of its relatively high abundance, 
and because the strongest resonance lines in He- and H-like O are in a relatively 
'clean' wavelength band. For reference, the Ly$\alpha$ transitions of \ion{C}{vi}, 
\ion{N}{vii}, \ion{O}{viii}, and \ion{Ne}{x} occur at 33.7360, 24.7810, 18.9689, 
and 12.1339 \AA, respectively (wavelengths of the $1s-2p_{1/2,3/2}$ doublet weighted 
with oscillator strength; \citet{johnson1985}. The He-like ions \ion{C}{v}, 
\ion{N}{vi}, \ion{O}{vii}, and \ion{Ne}{ix} have their strongest transition, 
the $n=1-2$ resonance line, at 40.2674, 28.7800, 21.6015, and 13.4473 \AA\ 
(\citealt{drake1988}; see also Table 1). 
Data on the higher order series members can be found in \citet{verner1996}. As far as 
the Fe L shell ions are concerned, the most likely transition to show up would be the 
strongest line in Ne-like \ion{Fe}{xvii}, $n=2p-3d$ $\lambda 15.014$ \AA. In addition, 
all lower ionisation stages of C, N, O, and Ne (with the exception of neutral Ne of course) 
can also absorb by $n=1-2$; the strongest of these transitions would be $1s-2p$ in 
\ion{O}{vi} at 22.019 \AA\ \citep{schmidt2004}. Likewise, the lower ionisation stages of 
Fe could in principle produce $n=2-3$ absorption.

The thermal widths of all these transitions will be very small, requiring resolving powers 
of order $R\sim 10\,000$ (C, N, O, Ne) for gas temperatures of order $10^6$ K to be resolved; for Fe, 
the requirement is even higher, by a factor $\sim 2$. As we will see, for practical reasons, 
these requirements exceed the current capabilities of astrophysical X-ray spectroscopy by a 
large factor. Due to the small Doppler broadening (ignoring turbulent velocity fields for now), 
the lines will rapidly saturate. For He- and H-like O resonance line absorption, saturation sets 
in at an equivalent width of order 1 m\AA\ (\citealt{kaastra2008}  - Chapter 9, this volume), or 
column densities of order a few times $10^{14}$ ions cm$^{-2}$. The challenge, therefore, for 
X-ray spectroscopy presented by the IGM is to detect small equivalent width, near-saturation 
lines that are unresolved.

\subsection{The baryon content of the WHIM as measured by UV and X-ray absorbers}
\label{The baryon content of the WHIM as measured by UV and X-ray absorbers}

One important result from absorption line measurements of the WHIM in the UV
is the observed number density of WHIM absorbers, usually expressed as 
${\mathrm d}N/{\mathrm d}z$, the number of absorbers per unit redshift. For instance, from
recent measurements with FUSE and HST/STIS one finds for \ion{O}{vi} absorbers and 
Broad Ly\,$\alpha$ absorbers at $z\approx0$ values of
${\mathrm d}N/{\mathrm d}z$(\ion{O}{vi}$)\approx 20$ and ${\mathrm d}N/{\mathrm d}z$(BLA$)\approx 30$ 
(see Sect.\,3.2).
Currently, the WHIM absorber density is only measurable in the UV,
since in the X-ray band both the observed number of WHIM absorption lines and
the available redshift path for WHIM observations is too small to derive
statistically significant values of ${\mathrm d}N/{\mathrm d}z$(\ion{O}{vii}) and ${\mathrm d}N/{\mathrm d}z$(\ion{O}{viii}).

A particularly interesting question now is, how the observed number density of 
high-ion lines or BLAs translates into an estimate of the cosmological
baryon mass density of the WHIM, $\Omega_{\mathrm b}$(WHIM). 
To obtain such an estimate of the baryon content of the WHIM from UV and X-ray
absorption measurements one has to consider two main steps. First, one needs to
transform the observed column densities of the high ions (e.g., \ion{O}{vi},
\ion{O}{vii}, \ion{O}{viii}) into a total gas column density by 
modelling the ionisation conditions in the gas. In a second step, one then
integrates over the total gas column densities of all observed WHIM absorbers 
along the given redshift path and from that derives $\Omega_{\mathrm b}$(WHIM) for 
a chosen cosmology. Throughout the paper we will assume 
a $\Lambda$CDM cosmology with $H_0=70$ km\,s$^{-1}$\,Mpc$^{-1}$,
$\Omega_{\Lambda}=0.7$, $\Omega_{\mathrm m}=0.3$, and $\Omega_{\mathrm b}=0.045$.
For the first step the uncertainty lies in the estimate of 
the ionisation fraction of hydrogen of the WHIM. For this, it is 
usually assumed that the WHIM is in collisional ionisation
equilibrium, but photoionisation and non-equilibrium conditions 
may play a significant role.
In the case of using metal ions such as \ion{O}{vi} 
the unknown oxygen abundance (O/H) of the gas introduces an additional
uncertainty (see below) for the estimate of $\Omega_{\mathrm b}$(WHIM).
For the second step, it is important to have a large enough
sample of WHIM absorption lines and a sufficient total redshift path
along {\it different} directions in order to handle statistical errors
and the problem of cosmic variance.
As mentioned earlier, these requirements currently are fulfilled
only for the UV absorbers.

The cosmological mass density $\Omega_{\mathrm b}$ of \ion{O}{vi} 
absorbers (and, similarly, for other high ions) 
in terms of the current critical density $\rho_{\rm c}$ can
be estimated by
\begin{equation}
\Omega_{\rm b}(\mbox{O\,{\scriptsize VI}})=\frac{\mu\,m_{\rm H}\,H_0}
{\rho_{\rm c}\,c}\,\sum_{ij}\,\frac{N(\mbox{O\,{\scriptsize VI}})_{ij}}
{f_{\mbox{O\,{\scriptsize VI}},ij}\,{\rm (O/H)}_{ij}\,\Delta X_j}.
\end{equation}
In this equation, $\mu=1.3$ is the mean molecular weight, 
$m_{\rm H}=1.673 \times 10^{-27}$ kg is the mass per hydrogen
atom, $H_0$ is the adopted local Hubble constant, 
and $\rho_{\rm c}=3H_0\,^2/8 \pi G$ is the current critical density.
The index $i$ denotes an individual high-ion absorption system
along a line of sight $j$. Each measured high-ion absorption
system $i$ is characterised
by its measured ion column density (e.g., $N$(\ion{O}{vi})$_{ij}$),
the ionisation fraction of the measured ion (e.g., $f_{\mbox{O\,{\scriptsize VI}},ij}$),
and the local abundance of the element measured compared to
hydrogen (e.g., the local oxygen-to-hydrogen ratio, by number).
Each line of sight $j$ has a
characteristic redshift range $\Delta z$
in which high-ion absorption may be detected.
The corresponding comoving path length $\Delta X$
available for the detection of WHIM high-ion absorbers
then is given by:
\begin{equation}
\Delta X_j=(1+z)^2\,[\Omega_{\Lambda}+\Omega_{\rm m}(1+z)^3]^{-0.5}\,\Delta z_j.
\end{equation}
In analogy, we can write for the cosmological mass density of the 
BLAs:
\begin{equation}
\Omega_{\rm b}{\rm (BLA)}=\frac{\mu\,m_{\rm H}\,H_0}
{\rho_{\rm c}\,c}\,\sum_{ij}\,\frac{N(\mbox{H\,{\scriptsize I}})_{ij}}
{f_{\mbox{H\,{\scriptsize I}},ij}\,\Delta X_j}.
\end{equation}

As can be easily seen, the advantage of using BLAs for deriving the WHIM mass density is
that the metallicity of the gas is unimportant for the determination
of $\Omega_{\mathrm b}$. The disadvantage is, however, that the ionisation
corrections are very large and uncertain, since they are determined indirectly
from the BLA line widths (see Sect.\,2.2.1).

\section{UV measurements of the WHIM}
\label{UV measurements of the WHIM}

\subsection{Past and present UV instruments}
\label{Past and present UV instruments}

\begin{figure}    
\begin{center}
\includegraphics[width=\hsize]{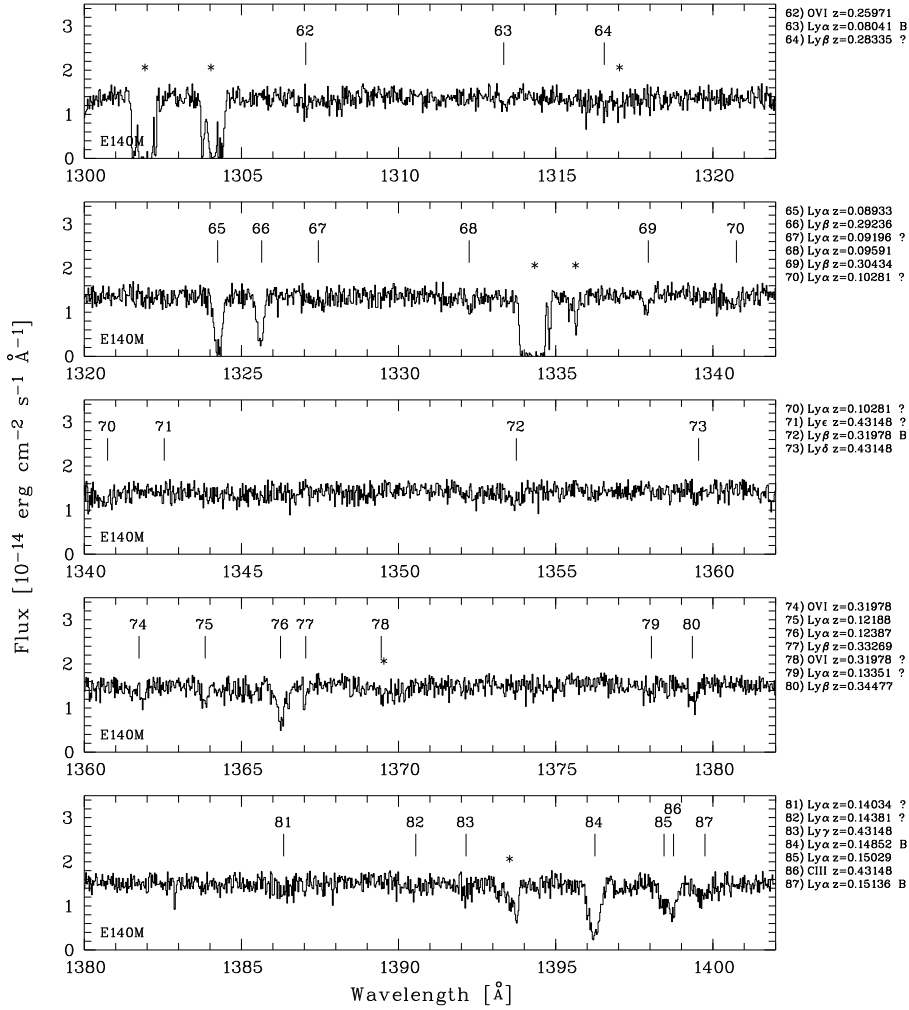}
\caption{
STIS spectrum of the quasar PG\,1259+593 in the wavelength range
between 1300 and 1400 \AA. 
Next to absorption from the local Ly\,$\alpha$ forest
and gas in the Milky Way there are several absorption features that
most likely are related to highly ionised gas in the WHIM. Absorption from five-times
ionised oxygen (\ion{O}{vi}) is observed at $z=0.25971$ and $z=0.31978$.
Broad \ion{H}{i} Ly\,$\alpha$ and Ly\,$\beta$ absorption is detected
at $z=0.08041$, $0.09196$, $0.10281$, $0.13351$, $0.14034$, $0.14381$,
$0.14852$, $0.15136$, and $z=0.31978$. From \protect\citet{richter2004}.
}
\label{fig:fig4}
\end{center}
\end{figure}

The first and second generations of space based UV spectrographs such
as {\it Copernicus} and the {\it International Ultraviolet Explorer}
(IUE) did not have sufficient sensitivity 
to systematically study intervening absorption in the intergalactic medium 
along a large number of sightlines. 
The early low- and intermediate resolution spectrographs installed 
on the {\it Hubble Space Telescope} (HST), namely the {\it Faint Object
Spectrograph} (FOS) and the {\it Goddard High Resolution Spectrograph}
(GHRS), were used to study the properties of the local Ly\,$\alpha$ forest
and intervening metal-line systems 
(e.g., \citealt{stocke1995}; \citealt{shull1998}).
While intervening \ion{O}{vi} absorption has been detected with 
these instruments (e.g., \citealt{tripp1998}), the concept of 
a warm-hot intergalactic gas phase was not really established
at that time.
With the implementation of the high-resolution 
capabilities of the {\it Space Telescope 
Imaging Spectrograph} (STIS) installed on HST 
the first systematic analyses of WHIM \ion{O}{vi} absorbers as 
significant low-redshift baryon reservoirs came out in 2000 (see \citealt{tripp2000}), 
thus relatively soon after the importance
of a shock-heated intergalactic gas phase was realised in cosmological 
simulations for the first time (e.g., \citealt{cen1999}; \citealt{dave2001}).
The STIS echelle spectrograph together with the E140M
grating provides a high spectral-resolution of $R\approx 45\,000$, 
corresponding to a velocity resolution of $\sim 7$ km\,s$^{-1}$
in the STIS E140M  wavelength band between $1150$ and $1730$ \AA\,
(e.g., \citealt{kimble1998}; \citealt{woodgate1998}). An example
for a STIS quasar spectrum with intervening hydrogen and metal-line
absorption is shown in Fig.~\ref{fig:fig4}.
Note that at the spectral resolution of the STIS E140M grating 
all intergalactic
absorption lines (i.e., hydrogen and metal lines) are fully resolved.
In 1999, the {\it Far Ultraviolet Spectroscopic Explorer} (FUSE) became available,
covering the wavelength range between $912$ and $1187$ \AA.
Equipped with a Rowland-type spectrograph providing a medium spectral resolution
of $R\approx 20\,000$ (FWHM$\sim20$ km\,s$^{-1}$) FUSE is able to observe
extragalactic UV background sources brighter than $V=16.5$ mag
with acceptable integration time and signal-to-noise (S/N) ratios
(for a description of FUSE see \citealt{moos2000}; \citealt{sahnow2000}).
With this resolution, FUSE is able to resolve the broader intergalactic
absorption from the \ion{H}{i} Lyman series, while most of the narrow metal-line
absorbers remain just unresolved. This is not a problem for \ion{O}{vi} WHIM studies
with FUSE, however, since the spectral resolution is very close to the actual 
line widths and the \ion{O}{vi} absorption usually is not saturated. 
FUSE complements the STIS instruments at lower wavelengths down to
the Lyman limit and consequently
combined FUSE and STIS spectra of $\sim 15$ low redshift QSOs and AGN have been used
to study the low-redshift WHIM via intervening \ion{O}{vi} and BLA absorption
(see \citet{tripp2007} and references therein). 
Unfortunately, since 2006/2007 both STIS
and FUSE are out of commission due to technical problems.

Fresh spectroscopic UV data from
WHIM absorption line studies will become available once the 
{\it Cosmic Origins Spectrograph} (COS) will be installed on HST during 
the next HST service mission (SM-4), which currently is scheduled for late
2008. COS will observe in the UV wavelength band between $1150$ and 
$3000$ \AA\, at medium resolution ($R\approx 20\,000$). COS has been
designed with maximum effective area as the primary constraint: it
provides more than an order of magnitude gain in sensitivity over previous 
HST instruments.
Due to its very high sensitivity, COS thus will be able to observe {\it hundreds} of
low- and intermediate redshift QSOs and AGN and thus will deliver an enormous data 
archive to study the properties of WHIM UV absorption lines systems in great detail
(see also \citealt{paerels2008} - Chapter 19, this volume).

\subsection{Intervening WHIM absorbers at low redshift}
\label{Intervening WHIM absorbers at low redshift}

\begin{figure}    
\begin{center}
\includegraphics[width=\hsize]{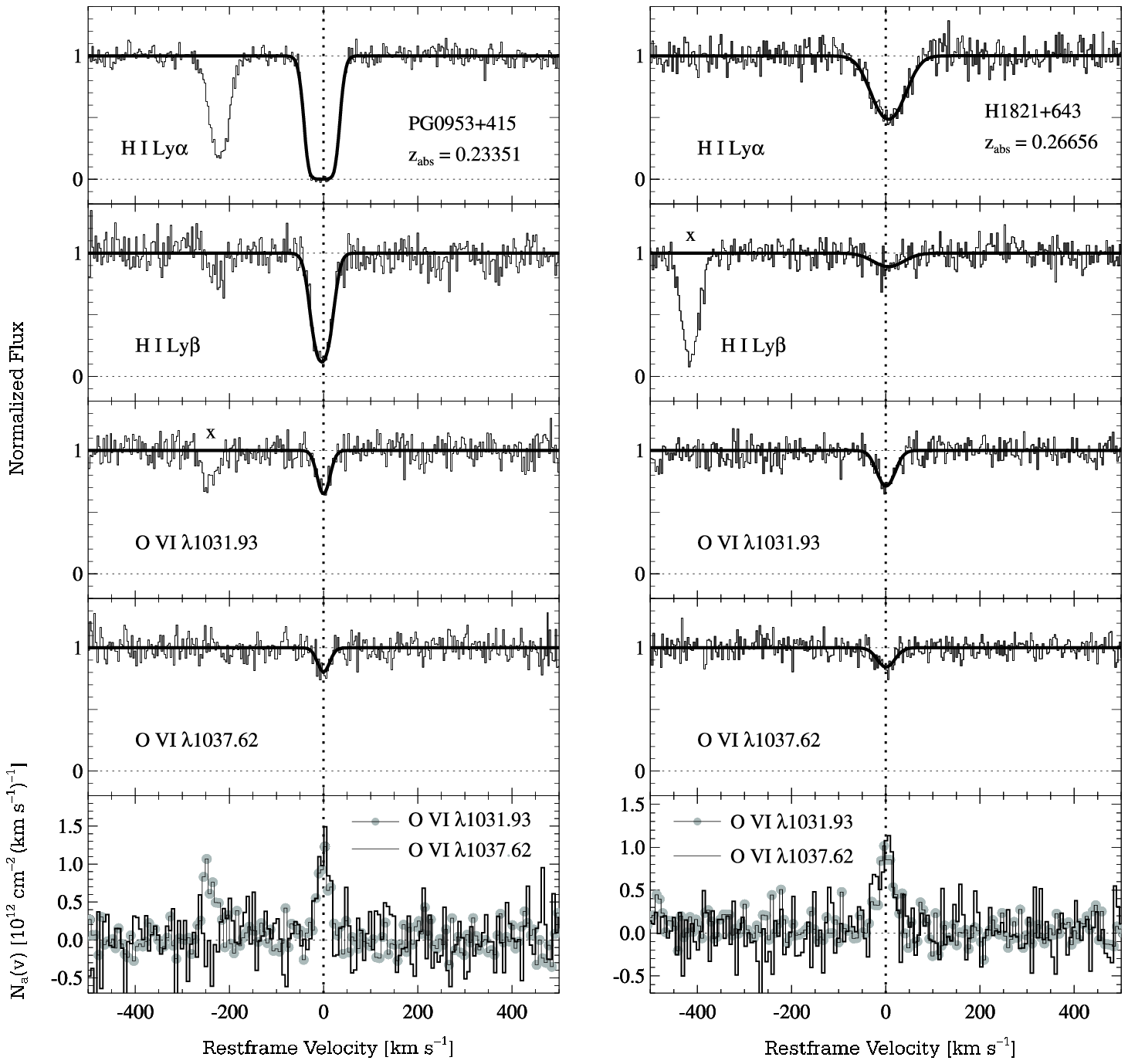}
\caption{
Examples for \ion{H}{i} and \ion{O}{vi} absorption in 
two absorption systems at $z=0.23351$ and $z=0.6656$
towards PG\,0953+415 and H\,1821+643, respectively,
plotted on a rest frame velocity scale
(observed with STIS).
Adapted from \protect\citet{tripp2007}.
}
\label{fig:fig5}
\end{center}
\end{figure}

We start with the \ion{O}{vi} absorbers which are believed
to trace the low-temperature tail of the 
WHIM  at $T<5\times 10^5$ K.
Up to now, more than 50 detections
of intervening \ion{O}{vi} absorbers at $z<0.5$
have been reported in the literature
(e.g., \citealt{tripp2000}; \citealt{oegerle2000};
\citealt{chen2000}; \citealt{savage2002};
\citealt{richter2004}; \citealt{sembach2004}; \citealt{savage2005};
\citealt{danforth2005}; \citealt{tripp2007}). All of these detections
are based on FUSE and STIS data.
Fig.~\ref{fig:fig5} shows two examples for intervening
\ion{O}{vi} absorption at $z=0.23351$ and $z=0.26656$ in the direction
of PG\,0953+415 and H\,1821+643, as observed with STIS.
The most recent compilation of low-redshift intervening \ion{O}{vi} absorbers
is that of \citet{tripp2007},
who have analysed 16 sightlines toward low-redshift QSOs 
observed with STIS and FUSE along a total redshift path
of $\Delta z\approx 3$.
They find a total of 53 intervening \ion{O}{vi} absorbers
(i.e., they are not within $5000$ km\,s$^{-1}$ of $z_{\rm QSO}$)
comprised of 78 individual absorption components\footnote{I.e., some of the \ion{O}{vi} systems have
velocity sub-structure}.
The measurements imply a number density of \ion{O}{vi}
absorbing systems per unit redshift of ${\mathrm d}N_{\mbox{O\,{\scriptsize VI}}}/{\mathrm d}z \approx 18\pm 3$
for equivalent widths $W_{\lambda}\geq30$ m\AA. The corresponding number
density of \ion{O}{vi} absorption {\it components} is 
${\mathrm d}N_{\mbox{O\,{\scriptsize VI}}}/{\mathrm d}z \approx 25\pm 3$. These values are 
slightly higher than what is found by
earlier analyses of smaller \ion{O}{vi} samples \citep{danforth2005},
but lie within the cited $2\sigma$ error ranges.
The discrepancy between the measured \ion{O}{vi} number densities 
probably is due to the different approaches of estimating the redshift
path $\Delta z$ along which the \ion{O}{vi} absorption takes place.
If one assumes that the gas is in a collisional ionisation equilibrium,
i.e., that $\sim 20$ percent of the oxygen is present
in the form of \ion{O}{vi} ($f_{\mbox{O\,{\scriptsize VI}}}\leq 0.2$), 
and further assumes that the mean oxygen abundance is $0.1$ Solar,
the measured number density of \ion{O}{vi} absorbers
corresponds to a cosmological mass density
of $\Omega_{\mathrm b}$(\ion{O}{vi})$\approx 0.0020-0.0030$ $h_{70}\,^{-1}$.
These values imply that intervening \ion{O}{vi} absorbers trace
$\sim 5-7$ percent of the total baryon mass in the local Universe.
For the interpretation of $\Omega_{\mathrm b}$(\ion{O}{vi})
it has to be noted
that \ion{O}{vi} absorption traces
collisionally ionised gas at
temperatures around $3 \times 10^5$ K (and also
low-density, photoionised gas at lower temperatures), but not the
million-degree gas phase which probably contains
the majority of the baryons in the WHIM.

The recent analysis of \citet{tripp2007} indicates, however,
that this rather simple conversion from measured \ion{O}{vi} 
column densities to $\Omega_{\mathrm b}$(\ion{O}{vi}) may not be justified 
in general, as the CIE assumption possibly breaks down 
for a considerable fraction of the \ion{O}{vi} systems. 
From the measured line widths of the \ion{H}{i} Ly\,$\alpha$
absorption that is associated with the \ion{O}{vi} Tripp et al.\,conclude
that $\sim 40$ percent of their \ion{O}{vi} systems belong to
cooler, photoionised gas with $T<10^5$ K, possibly not at all
associated with shock-heated warm-hot gas. In addition, about half of the
intervening \ion{O}{vi} absorbers arise in rather complex, multi-phase
systems that can accommodate hot gas at relatively low metallicity.
It thus appears that -- without having additional information about the physical 
conditions in each \ion{O}{vi} absorber -- the estimate of the 
baryon budget in intervening \ion{O}{vi} systems is afflicted 
with rather large systematic uncertainties.

In high-column density \ion{O}{vi} systems at redshifts $z>0.18$, 
such desired additional information may be 
provided by the presence or absence of \ion{Ne}{viii} (see Sect.\,2.2.1),
which in CIE traces gas at $T\sim 7 \times 10^5$ K.
Toward the quasar PG\,1259+593 \citet{richter2004} have reported a
tentative detection of \ion{Ne}{viii} absorption
at $\sim 2 \sigma$ significance in an \ion{O}{vi} absorber
at $z\approx 0.25$. The first secure detection 
of intervening \ion{Ne}{viii} absorption 
(at $\sim 4 \sigma$ significance) was presented by
\citet{savage2005} in a multi-phase \ion{O}{vi} absorption system
at $z\approx 0.21$ in the direction of the quasar HE\,0226$-$4110.
The latter authors show that in this particular absorber the high-ion ratio
\ion{Ne}{viii}/\ion{O}{vi}$=0.33$ is in agreement with gas in CIE
at temperature of $T\sim 5\times 10^5$ K.
With future high S/N absorption line data of low-redshift QSOs
(as will be provided by COS) it is expected 
that the number of detections of WHIM \ion{Ne}{viii} absorbers 
will increase substantially, so that an important new diagnostic
will become available for the analysis of high-ion absorbers.

One other key aspect in understanding the distribution and nature
of intervening \ion{O}{vi} systems concerns their relation to the 
large-scale distribution of galaxies. 
Combining FUSE data of 37 \ion{O}{vi} absorbers with a database
of more than a million galaxy positions and redshifts,
\citet{stocke2006} find that all of these \ion{O}{vi} systems
lie within $800\,h_{70}\,^{-1}$ kpc of the nearest galaxy.
These results suggest that \ion{O}{vi} systems preferentially
arise in the immediate circumgalactic environment and extended halos
of galaxies, where the metallicity of the gas is expected to be
relatively high compared to regions far away from galactic
structures.
Some very local analogs of intervening \ion{O}{vi} systems thus may be 
the \ion{O}{vi} high-velocity clouds in the Local Group 
that are discussed in the next subsection.
Due to apparent strong connection between intervening 
\ion{O}{vi} systems and galactic structures 
and a resulting galaxy/metallicity bias problem it is 
of great interest to consider other tracers
of warm-hot gas, which are independent of the metallicity of the gas.
The broad hydrogen Ly$\alpha$ absorbers -- as will be 
discussed in the following -- therefore represent 
an important alternative for studying the WHIM at low redshift.

\begin{figure}    
\begin{center}
\includegraphics[width=\hsize]{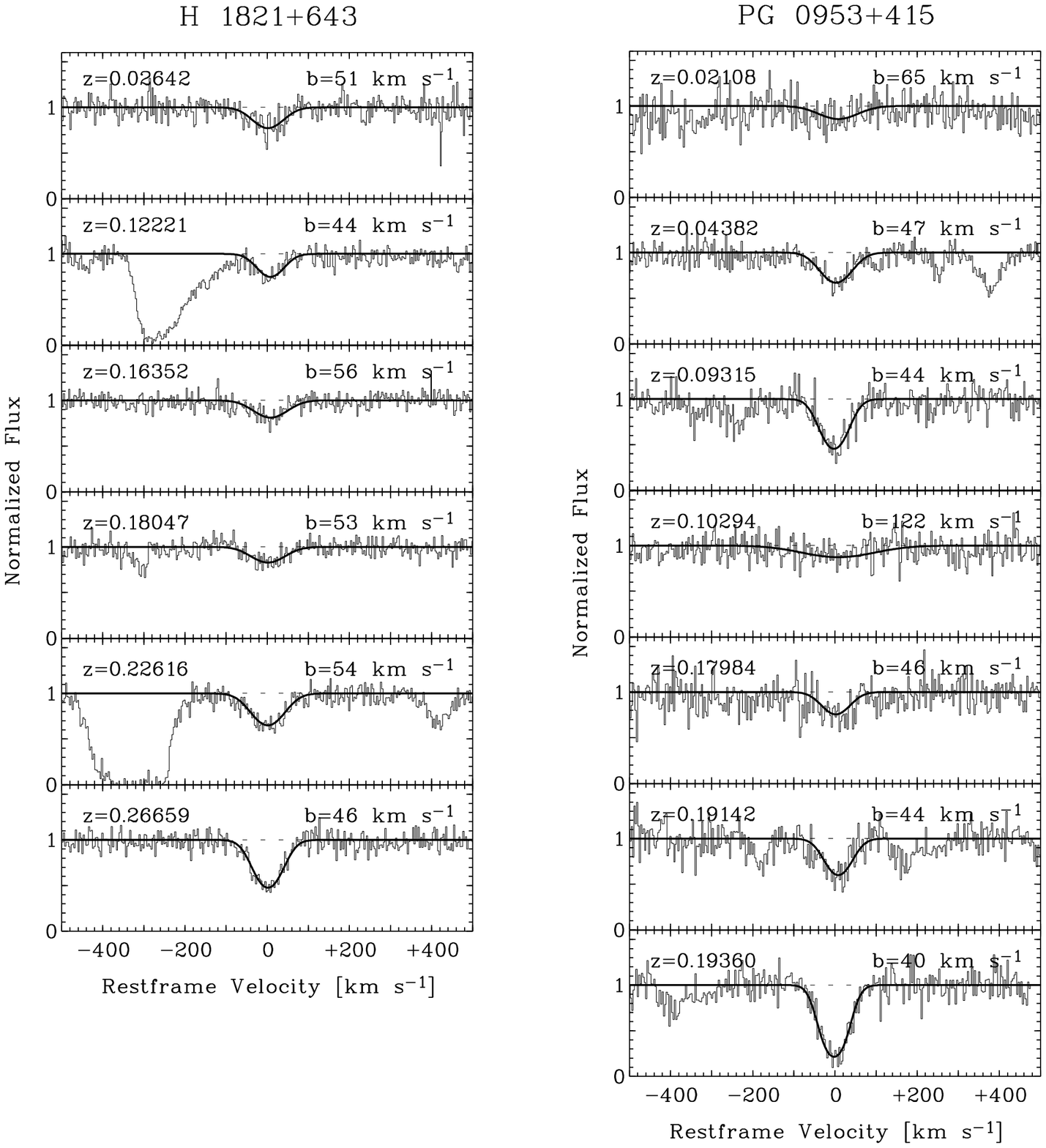}
\caption{
Broad Lyman $\alpha$ absorbers towards the
quasars H\,1821+643 and PG\,0953+415 (STIS observations), plotted on
a rest frame velocity scale.
If thermal line broadening dominates the width of the
absorption, these systems trace the WHIM at temperatures
between $10^5$ and $10^6$ K. From \protect\citet{richter2006a}.
}
\label{fig:fig6}
\end{center}
\end{figure}

As described in Sect.\,2.2.1, BLAs represent \ion{H}{i} Ly$\alpha$
absorbers with large Doppler parameters $b>40$ km\,s$^{-1}$.
If thermal line broadening dominates the width of the
absorption, these systems trace the WHIM at temperatures
between $10^5$ and $10^6$ K, typically (note that for most systems with 
$T>10^6$ K BLAs are both too broad and too shallow to be 
unambiguously identified with the limitations of current UV spectrographs).
The existence of \ion{H}{i} Ly$\alpha$ absorbers with 
relatively large line widths has been occasionally reported in 
earlier absorption-line studies of the local intergalactic 
medium (e.g., \citealt{tripp2001}; \citealt{bowen2002}).
Motivated by the rather frequent occurrence of broad absorbers 
along QSO sightlines with relatively large redshift paths, 
the first systematic analyses of BLAs in STIS low-$z$
data were carried out by \citet{richter2004} and
\citet{sembach2004}. \citet{richter2006a} have 
inspected four sightlines observed with STIS
towards the quasars PG\,1259+593
($z_{\rm em}=0.478$), PG\,1116+215 ($z_{\rm em}=0.176$),
H\,1821+643 ($z_{\rm em}=0.297$), and PG\,0953+415 ($z_{\rm em}=0.239$)
for the presence of
BLAs and they identified a number of good candidates.
Their study implies a BLA
number density per unit redshift
of ${\mathrm d}N_{\rm BLA}/{\mathrm d}z \approx 22-53$ for Doppler parameters
$b\geq40$ km\,s$^{-1}$ and above a sensitivity limit of
log ($N$(cm$^{-2})/b($km\,s$^{-1}))\geq 11.3$.
The large range for ${\mathrm d}N_{\rm BLA}/{\mathrm d}z$ partly is due to the uncertainty
about defining reliable selection criteria for
separating spurious cases from good broad Ly$\alpha$ candidates
(see discussions in \citealt{richter2004,richter2006a} and \citealt{sembach2004}).
Transforming the number density ${\mathrm d}N_{\rm BLA}/{\mathrm d}z$ into a cosmological
baryonic mass density, \citet{richter2006a} 
obtains $\Omega_{\mathrm b}$(BLA)$\geq 0.0027\,h_{70}\,^{-1}$.
This limit is about 6 percent of the total baryonic mass density
in the Universe expected from the current cosmological models (see above),
and is comparable with the value derived for the intervening \ion{O}{vi} absorbers
(see above). Examples for several BLAs in the STIS
spectrum of the quasar H\,1821+643 are shown in Fig.~\ref{fig:fig6}.

More recently, \citet{lehner2007} have analysed BLAs in low-redshift
STIS spectra along seven sightlines. They find a BLA number density
of ${\mathrm d}N_{\rm BLA}/{\mathrm d}z=30\pm4$ for $b=40-150$ km\,s$^{-1}$ and
log $N$(\ion{H}{i}$)>13.2$ for the redshift range $z=0-0.4$. They
conclude that BLAs host at least 20 percent of the baryons in the
local Universe, while the photoionised Ly$\alpha$ forest, which
produces a large number of narrow Ly$\alpha$ absorbers (NLAs),
contributes with $\sim 30$ percent to the total baryon budget.
In addition, \citet{prause2007} have investigated 
the properties of BLAs at intermediate redshifts ($z=0.9-1.9$)
along five other quasars using STIS high- and intermediate-resolution
data. They find a number density of reliably detected BLA candidates
of ${\mathrm d}N_{\rm BLA}/{\mathrm d}z\approx14$ 
and obtain a lower limit of the contribution of BLAs to the total
baryon budget of $\sim 2$ percent in this redshift range. The frequency
and baryon content of BLAs at intermediate redshifts obviously 
is lower than at $z=0$, indicating that at intermediate redshifts
shock-heating of the intergalactic gas from the infall 
in large-scale filaments is not yet very efficient.
This is in line with the predictions from 
cosmological simulations.

\subsection{The Milky Way halo and Local Group gas}
\label{The Milky Way halo and Local Group gas}

One primary goal of the FUSE mission was to constrain the 
distribution and kinematics of hot gas in the thick
disk and lower halo of the Milky Way by studying the properties
of Galactic \ion{O}{vi} absorption systems at radial
velocities $|v_{\rm LSR}|\leq 100$ km\,s$^{-1}$
(\citealt{savage2000}; \citealt{savage2003}; \citealt{wakker2003}).
However, as the FUSE data unveil, 
\ion{O}{vi} absorption associated with Milky Way gas
is observed not only at low velocities
but also at $|v_{\rm LSR}|>100$ km\,s$^{-1}$ \citep{sembach2003}.
The topic of cool and hot gas in the halo of the Milky Way
recently has been reviewed by \citet{richter2006c}.
These detections imply that next to the Milky Way's hot "atmosphere"
(i.e., the Galactic Corona)
individual pockets of hot gas exist that move
with high velocities through
the circumgalactic environment of the Milky Way.
Such high-velocity \ion{O}{vi} absorbers may contain
a substantial fraction of the baryonic matter in the
Local Group in the form of warm-hot gas and thus
-- as discussed in the previous subsection --
possibly represent the local counterparts of some of the 
intervening \ion{O}{vi} absorbers observed towards 
low-redshift QSOs.

From their FUSE survey of high-velocity \ion{O}{vi} absorption
\citet{sembach2003} find that probably more than 60 percent
of the sky at high velocities is covered by ionised hydrogen
(associated with the \ion{O}{vi} absorbing gas) above
a column density level of log $N$(\ion{H}{ii})$=18$, assuming
a metallicity of the gas of $0.2$ Solar.
Some of the high-velocity \ion{O}{vi} detected with FUSE
appears to be associated with known high-velocity \ion{H}{i} 21~cm
structures (e.g., the high-velocity clouds complex A,
complex C, the Magellanic Stream, and the Outer Arm).
Other high-velocity \ion{O}{vi} features, however, have no counterparts
in \ion{H}{i} 21~cm emission. The high radial velocities for most of these
\ion{O}{vi} absorbers are incompatible with those expected
for the hot coronal gas (even if the coronal gas motion is
decoupled from the underlying rotating disk). A transformation
from the Local Standard of Rest to the Galactic Standard of
Rest and the Local Group Standard of Rest velocity reference frames
reduces the dispersion  around the mean of the high-velocity
\ion{O}{vi} centroids (\citealt{sembach2003}; \citealt{nicastro2003}). This can be
interpreted as evidence that {\it some} of the \ion{O}{vi} high-velocity
absorbers are intergalactic clouds in the Local Group
rather than clouds directly associated with the Milky Way.
However, it is extremely difficult to discriminate between a Local
Group explanation and a distant Galactic explanation for these
absorbers.
The presence of intergalactic \ion{O}{vi} absorbing gas
in the Local Group is in line with
theoretical models that predict that there should be a large
reservoir of hot gas left over from the formation
of the Local Group (see, e.g., \citealt{cen1999}).

It is unlikely that the high-velocity \ion{O}{vi} is
produced by photoionisation. Probably, the gas
is collisionally ionised at temperatures of several
$10^5$ K. The \ion{O}{vi} then may be produced
in the turbulent interface regions between very hot
($T>10^6$ K) gas in an extended Galactic Corona
and the cooler gas clouds that are moving through
this hot medium (see \citealt{sembach2003}).
Evidence for the existence of such
interfaces also comes from the comparison of absorption
lines from neutral and weakly ionised species with
absorption from high ions
like \ion{O}{vi} \citep{fox2004}.

\section{X-ray measurements of the WHIM}
\label{X-ray measurements of the WHIM}

\subsection{Past and present X-ray instruments}
\label{Past and present X-ray instruments}

With the advent of the {\it Chandra} and {\it XMM-Newton} observatories, 
high resolution X-ray spectroscopy of a wide variety of cosmic sources became 
feasible for the first time.
Among the possible results most eagerly speculated upon was the detection 
of intergalactic absorption lines from highly ionised metals in the 
continuum spectra of bright extragalactic sources. After all, one of 
the most striking results from the {\it Einstein} observatory had been 
the detection of a very significant broad absorption feature at $\sim 600$ 
eV in the spectrum of PKS\,2155$-$304 with the Objective Grating Spectrometer 
\citep{canizares1984}. Ironically, if interpreted as intergalactic 
H-like O Ly$\alpha$ absorption, redshifted and broadened by the expansion 
of the Universe, the strength of the feature implied the presence of a highly 
ionised IGM of near-critical density, a possibility that has of course 
definitively been discounted since then.

The High Energy Transmission Grating Spectrometer (HETGS; \citealt{canizares2005}) 
and the Low Energy Transmission Grating Spectrometer (LETGS; \citealt{brinkman2000}) 
on {\it Chandra}, and the Reflection Grating Spectrometer (RGS) on 
{\it XMM-Newton} \citep{denherder2001} were the first instruments to provide 
sensitivity to weak interstellar and intergalactic X-ray absorption lines. 
The 'traditional' ionisation detectors (proportional counters, CCD's) do not have 
sufficient energy resolution for this application.
But the angular resolution provided by an X-ray telescope can be used to produce 
a high resolution spectrum, by the use of diffracting elements. Laboratory X-ray 
spectroscopy is typically performed with crystal diffraction spectrometers, and 
the use of crystal spectrometers for general use in astrophysics was pioneered 
on the {\it Einstein} observatory (e.g. \citealt{canizares1979}). The Focal Plane 
Crystal Spectrometer indeed detected the first ever narrow X-ray absorption line in a 
cosmic source, the $1s-2p$ absorption by neutral oxygen in the interstellar medium 
towards the Crab \citep{canizares1984}.
Previous grating spectrometers (the Objective Grating Spectrometer on {\it Einstein} 
and the two Transmission Grating Spectrometers on {\it EXOSAT}) had only limited 
resolution and sensitivity. But there is no fundamental limit to the resolution of 
a diffraction grating spectrometer, and the high angular resolution of the {\it Chandra} 
telescope has allowed for high resolution spectroscopy using transmission gratings. 
The focusing optics on {\it XMM-Newton} have more modest angular resolution, but they 
are used with a fixed array of grazing incidence reflection gratings, which produce 
very large dispersion angles and thus high spectral resolution.
The HETGS provides a spectral resolution of $\Delta\lambda = 0.0125$ \AA\ over the 
$\approx 1.5-15$ \AA\ band with the high line density grating, and
$\Delta\lambda = 0.025$ \AA\ over the $2-20$ \AA\ band with the medium line density 
grating. The LETGS has  $\Delta\lambda = 0.05$ \AA\ over the $\approx 2-170$ \AA\ band, 
while the RGS has  $\Delta\lambda = 0.06$ \AA\ over the $5-38$ \AA\ band. These numbers 
translate to resolving powers of $R = 400-1500$ in the O K band, and with a 
sufficiently bright continuum source, one should be able to detect equivalent widths of 
order  $0.1-0.5$~eV ($5-20$~m\AA), or below in spectra with very high signal-to-noise. 
The predictions for H- and He-like O resonance absorption line strengths are 
generally smaller than these thresholds, but not grossly so, and so a search for 
intergalactic O was initiated early on.
Given that the current spectrometers are not expected to resolve the absorption lines, 
the only freedom we have to increase the sensitivity of the search is to increase the 
signal-to-noise ratio in the continuum, and it becomes crucial to find suitable, very 
bright sources, at redshifts that are large enough that there is a reasonable a priori 
probability of finding a filament with detectable line absorption. 

As we will see when we discuss the results of the observational searches for X-ray 
absorption lines, the problem is made considerably more difficult by the very 
sparseness of the expected absorption signature. Frequently, when absorption features 
of marginal statistical significance are detected in astrophysical data, plausibility 
is greatly enhanced by simple, unique spectroscopic arguments. For instance, for all 
plausible parameter configurations, an absorbing gas cloud detected in \ion{N}{v} 
should also produce detectable \ion{C}{iv} absorption; or, both members of a doublet 
should appear in the correct strength ratio if unsaturated. In the early stages of 
Ly$\alpha$ forest astrophysics, it was arguments of this type, rather than the crossing 
of formal statistical detection thresholds alone, that guided the field 
(e.g., \citealt{lynds1971}). But for the 'X-ray Forest' absorption, we expect a very 
different situation. The detailed simulations confirm what simple analytical arguments 
had suggested: in most cases, Intergalactic absorption systems that are in principle 
detectable with current or planned X-ray instrumentation will show just a single 
absorption line, usually the \ion{O}{vii} $n=1-2$ resonance line, at an unknown redshift. 
When assessing the possibility that a given apparent absorption feature is 'real', 
one has to allow for the number of independent trial redshifts (very roughly given by 
the width of the wavelength band surveyed, divided by the nominal spectral resolution 
of the spectrometer), and with a wide band and a high spectral resolution, this tends 
to dramatically reduce the significance of even fairly impressive apparent absorption 
dips. For instance, an apparent absorption line detected at a formal '3$\sigma$ significance' 
(or $p = 0.0015$ a priori probability for a negative deviation this large to occur due to 
statistical fluctuation) with {\it Chandra} LETGS in the $21.6-23.0$ \AA\ band pales to 
'1.7 $\sigma$' if we assume it is the \ion{O}{vii} resonance line in the redshift 
range $z = 0-0.065$; with $N \approx 30$ independent trials, the chances of not 
seeing a $3\sigma$ excursion are $(1-p)^N = 0.956$, or: one will see such a feature 
one in twenty times if one tries this experiment (we are assuming a Gaussian distribution 
of fluctuations here). If we allow for confusion with \ion{O}{viii} Ly$\alpha$ at higher 
redshift, or even other transitions, the significance is even further reduced. And the larger 
the number of sources surveyed, the larger the probability of false alarm. Clearly, more
reliable statistics on intervening X-ray absorbers and detections at higher
significance are desired, but the required high-quality data will not be 
available until the next-generation X-ray facilities such as {\it XEUS} and 
{\it Constellation X} are installed
(see \citealt{paerels2008}  - Chapter 19, this volume).

Nevertheless, even with these odds, the above discussed high-ion measurements 
are important observations to do with the
currently available instruments {\it Chandra} and {\it XMM-Newton}.
Given the predicted strengths of the absorption lines (e.g., \citealt{chen2003}; see also 
Sect.\,5.1), attention has naturally focused on a handful of very bright 
BL Lac- and similar sources. Below, we discuss the results of the searches. Note that the 
subject has recently also been reviewed by \citet{bregman2007}.

\subsection{Intervening WHIM absorbers at low redshift}
\label{Intervening WHIM absorbers at low redshift X-ray}

The first attempt at detecting redshifted X-ray O absorption lines was performed 
by \citet{mathur2003} with a dedicated deep observation (470 ks) with the 
{\it Chandra} LETGS of the quasar H\,1821+643, which has several confirmed 
intervening \ion{O}{vi} absorbers. No significant X-ray absorption 
lines were found at the redshifts of the \ion{O}{vi} systems, but this was not 
really surprising in view of the modest signal to noise ratio in the X-ray continuum.
Since it requires very bright continua to detect the weak absorption, it is also 
not surprising that the number of suitable extragalactic sources is severely limited. 
Early observations of a sample of these
(e.g. S5\,0836+710, PKS\,2149$-$306; \citealt{fang2001b}; PKS\,2155$-$304, \citealt{fang2002}) 
produced no convincing detections. Nicastro and his colleagues then embarked on a 
campaign to observe Mrk\,421 during its periodic X-ray outbursts, when its X-ray flux rises 
by an order of magnitude (e.g. \citealt{nicastro2005}). The net result of this has been the accumulation 
of a very deep spectrum with the {\it Chandra} LETGS, with a total of more than 7 million 
continuum counts, in about 1000 resolution elements. \citet{nicastroetal2005} have claimed 
evidence for the detection of two intervening absorption systems in these data, 
at $z = 0.011$ and $z = 0.027$. But the spectrum of the same source observed 
with the {\it XMM-Newton} RGS does not show these absorption lines \citep{rasmussen2007}, 
despite higher signal-to-noise and comparable spectral resolution (Mrk 421 is 
observed by {\it XMM-Newton} for calibration purposes, and by late 2006, 
more than 1 Ms exposure had been accumulated). \citet{kaastra2006} have 
reanalysed the {\it Chandra} LETGS data, and find no significant absorption.
Other sources, less bright but with larger redshifts, have been observed 
(see for instance \citet{steenbrugge2006} for observations of 
1ES\,1028+511 at $z=0.361$), but to date no convincing evidence for 
intervening absorption has materialised.

Observations have been conducted to try and detect the absorption 
by intergalactic gas presumably associated with known locations of 
cosmic overdensity, centred on massive clusters.
\citet{fujimoto2004} attempted to detect absorption in the quasar LBQS 1228+1116, 
located behind the Virgo cluster. An {\it XMM-Newton} RGS spectrum revealed a
marginal feature at the (Virgo) redshifted position of \ion{O}{viii} Ly$\alpha$, 
but only at the $\sim 95$\% confidence level. Likewise,
\citet{takei2007} took advantage of the location of X Comae behind the 
Coma cluster to try and detect absorption from Coma or its surroundings, 
but no convincing, strong absorption lines were detected in a deep 
observation with {\it XMM-Newton} RGS. The parallel CCD imaging data obtained 
with EPIC show weak evidence for \ion{Ne}{ix} $n=1-2$ line emission at 
the redshift of Coma, which, if real, would most likely be associated with 
WHIM gas around the cluster, seen in projection (the cluster virial temperature 
is too high for \ion{Ne}{ix}). In practice, the absence of very bright point 
sources behind clusters, which makes absorption studies difficult, and the 
bright foregrounds in emission, will probably make this approach to detecting 
and characterizing the WHIM not much easier than the random line-of-sight searches.

The conclusion from the search for intergalactic X-ray absorption is 
that there is no convincing, clear detection
for intervening absorption. This is, in retrospect, not that surprising, 
given the sensitivity of the current X-ray spectrometers,
the abundance of suitably bright and sufficiently distant continuum 
sources, and the predicted properties of the WHIM.

\subsection{The Milky Way halo and Local Group gas}
\label{The Milky Way halo and Local Group gas X-ray}

\begin{figure}    
\begin{center}
\includegraphics[width=\hsize]{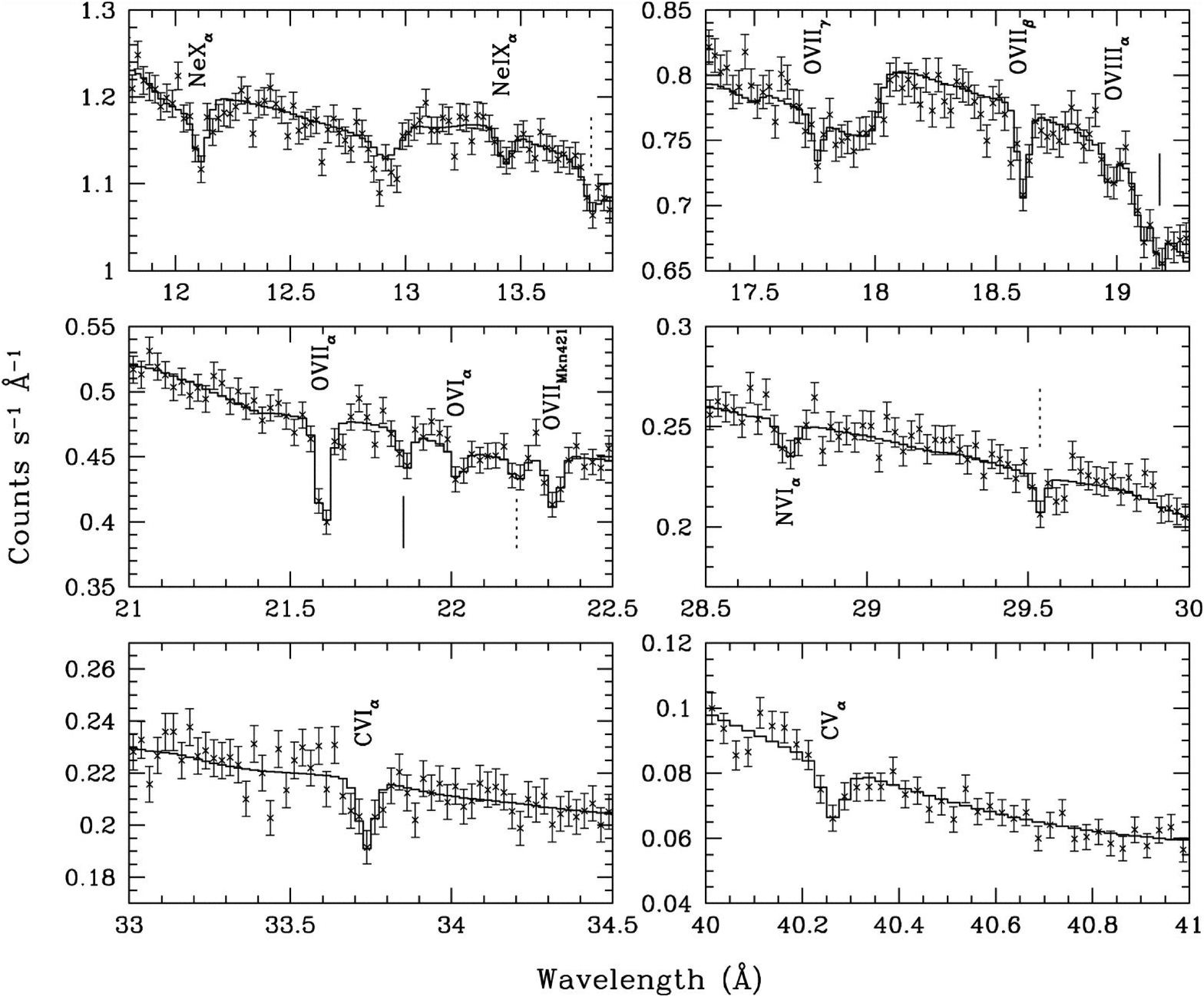}

\caption{{\it Chandra} LETGS spectrum of Mrk\,421. Crosses are the data, 
the solid line is a model. The labels identify $z \approx 0$ absorption 
lines in Ne, O, and C. The vertical tick marks indicate the locations of 
possible intergalactic absorption lines. From \protect\citet{williams2005}.
}
\label{fig:fig7}
\end{center}
\end{figure}

The first positive result of the analysis of bright continuum spectra was 
the detection of \ion{O}{vii} and \ion{O}{viii} $n=1-2$ resonance line 
absorption at redshift zero. \citet{nicastro2002} first identified the 
resonance lines in the {\it Chandra} LETGS spectrum of PKS\,2155$-$304 (\ion{O}{vii} 
$n=1-2$, $n=1-3$, \ion{O}{viii} Ly$\alpha$, \ion{Ne}{ix} $n=1-2$). \citet{rasmussen2003} 
detected resonance absorption in the {\it XMM-Newton} RGS spectra of 3C\,273, Mrk\,421, 
and PKS\,2155$-$304. Since then, at least \ion{O}{vii} $n=1-2$ has been detected in 
effectively all sufficiently bright continuum sources, both with {\it Chandra} 
and {\it XMM-Newton}; a recent compilation appears in \citet{fang2006}. 
Portions of a deeper spectrum that shows the zero redshift absorption are 
shown in Fig.~\ref{fig:fig7}.

\citet{nicastro2002} initially interpreted the absorption as arising in an extended intergalactic filament. The argument that drives this interpretation is based on the
assumption that \ion{O}{vi}, \ion{O}{vii}, and \ion{O}{viii} are all located in a single phase of the absorbing gas. The simultaneous appearance of finite amounts of
\ion{O}{vi} and \ion{O}{viii} only occurs in photoionised gas, not in gas in collisional ionisation equilibrium, and this requires that the gas has very low density
(the photoionisation is produced by the local X-ray background radiation field; for a measured ionisation parameter, the known intensity of the ionising field fixes the
gas density). The measured ionisation balance then implies a length scale on the order of $l \sim 10\ (Z_{0.1})^{-1}$ Mpc, where $Z_{0.1}$ is the metallicity in units
0.1 Solar. This is a very large length, and even for $0.3\ Z_{\odot}$ metallicity, the structure still would not fit in the Local Group (and it is unlikely to have this
high a metallicity if it were larger than the Local Group). In fact, the absorption lines should have been marginally resolved in this case, if the structure expands
with a fair fraction of the expansion of the Universe.

\citet{rasmussen2003} constrained the properties of the absorbing gas by dropping the \ion{O}{vi}, and by taking into account the intensity of the
diffuse \ion{O}{vii} and \ion{O}{viii} line emission as measured by the Wisconsin/Goddard rocket-borne Quantum X-ray Calorimeter (XQC) experiment 
\citep{mccammon2002}. The cooling timescale of \ion{O}{vi}-bearing gas is much smaller than that of gas with the higher ionisation stages of O, and this justifies the
assumption that \ion{O}{vi} is located in a different, transient phase of the gas. With only \ion{O}{vii} and \ion{O}{viii}, the medium can be denser and more compact,
and be in collisional ionisation equilibrium. Treating the measured \ion{O}{vii} emission line intensity as an upper limit to the emission from a uniform medium, and
constraining the ionisation balance from the measured ratio of \ion{O}{vii} and \ion{O}{viii} column densities in the lines of sight to Mrk\,421 and PKS\,2155$-$304,
Rasmussen et al. derived an upper limit on the density of the medium of $n_{\mathrm e} \lsim 2 \times 10^{-4}$~cm$^{-3}$ and a length scale $l \gsim 100$~kpc.

\citet{bregman2007} favours a different solution, with a lower electron temperature and hence a higher \ion{O}{vii} ion fraction. If one assumes Solar abundance and
sets the \ion{O}{vii} fraction to 0.5, the characteristic density becomes $n_e \sim 10^{-3}$~cm$^{-3}$, and the length scale ($l \sim 20$~kpc) suggests a hot Galactic
halo, rather than a Local Group intragroup medium.

Arguments for both type of solution (a small compact halo and a more tenuous Local Group medium) can be given. The most direct of these is a measurement of the
\ion{O}{vii} line absorption towards the LMC by \citet{wang2005} in the spectrum of the X-ray binary LMC X-3, which indicates that a major fraction of the
\ion{O}{vii} column in that direction is in fact in front of the LMC. \citet{bregman2007} points out that the distribution of column densities of highly ionised
O on the sky is not strongly correlated with the likely projected mass distribution of the Local Group, and that the measured velocity centroid of the absorption lines
appears characteristic of Milky Way gas, rather than Local Group gas. On the other hand, a direct measurement of the Doppler broadening of the \ion{O}{vii} gas, from
the curve of growth of the $n=1-2$ and $n=1-3$ absorption lines in the spectrum of Mrk\,421 and PKS\,2155$-$304 (\citealt{williams2005},
\citealt{williams2007}), indicates an ion temperature of $T_{\rm i} \approx 10^{6.0-6.3}$~K (Mrk\,421) and $T_i \approx 10^{6.2-6.4}$~K (PKS\,2155$-$304), and these
values favour the low-density, Local Group solution. Regardless, the prospect of directly observing hot gas expelled from the Galactic disk, or measuring the virial
temperature of the Galaxy and/or the Local Group is exciting enough to warrant further attention to redshift zero absorption and emission.

Finally, the spectrum of Mrk\,421 shows the expected $z = 0$ innershell \ion{O}{vi} absorption, at 22.019~\AA, both with {\it Chandra} and with {\it XMM-Newton}. There
has been some confusion regarding an apparent discrepancy between the \ion{O}{vi} column densities derived from the FUV and from the X-ray absorption lines, in the
sense that the X-ray column appeared to be significantly larger than the FUV column \citep{williams2005}. Proposed physical explanations for this
effect involve a depletion of the lower level of the FUV transitions ($1s^2 2s$) in favour of (at least) $1s^2 2p$, which weakens the $\lambda\lambda 1032, 1038$ \AA\
absorption but does not affect the $1s-2p$ X-ray absorption. However, it requires very high densities to maintain a finite excited state population, and, more directly,
the measured wavelength of the X-ray line is actually not consistent with the wavelength calculated for $1s-2p$ in excited \ion{O}{vi}, off by about $0.03-0.05$ \AA, on
the order of a full resolution element of both the {\it Chandra} LETGS and the {\it XMM-Newton} RGS (Raassen 2007, private communication). The conclusion is that the
discrepancy is due to an authentic statistical fluctuation in the X-ray spectrum -- or, more ironically, to the presence of a weak, slightly redshifted \ion{O}{vii}
absorption line.

\section{Additional aspects}\label{Additional aspects}

\subsection{Results from numerical simulations}\label{Results from numerical simulations}

\begin{figure}    
\begin{center}
\includegraphics[width=0.75\hsize]{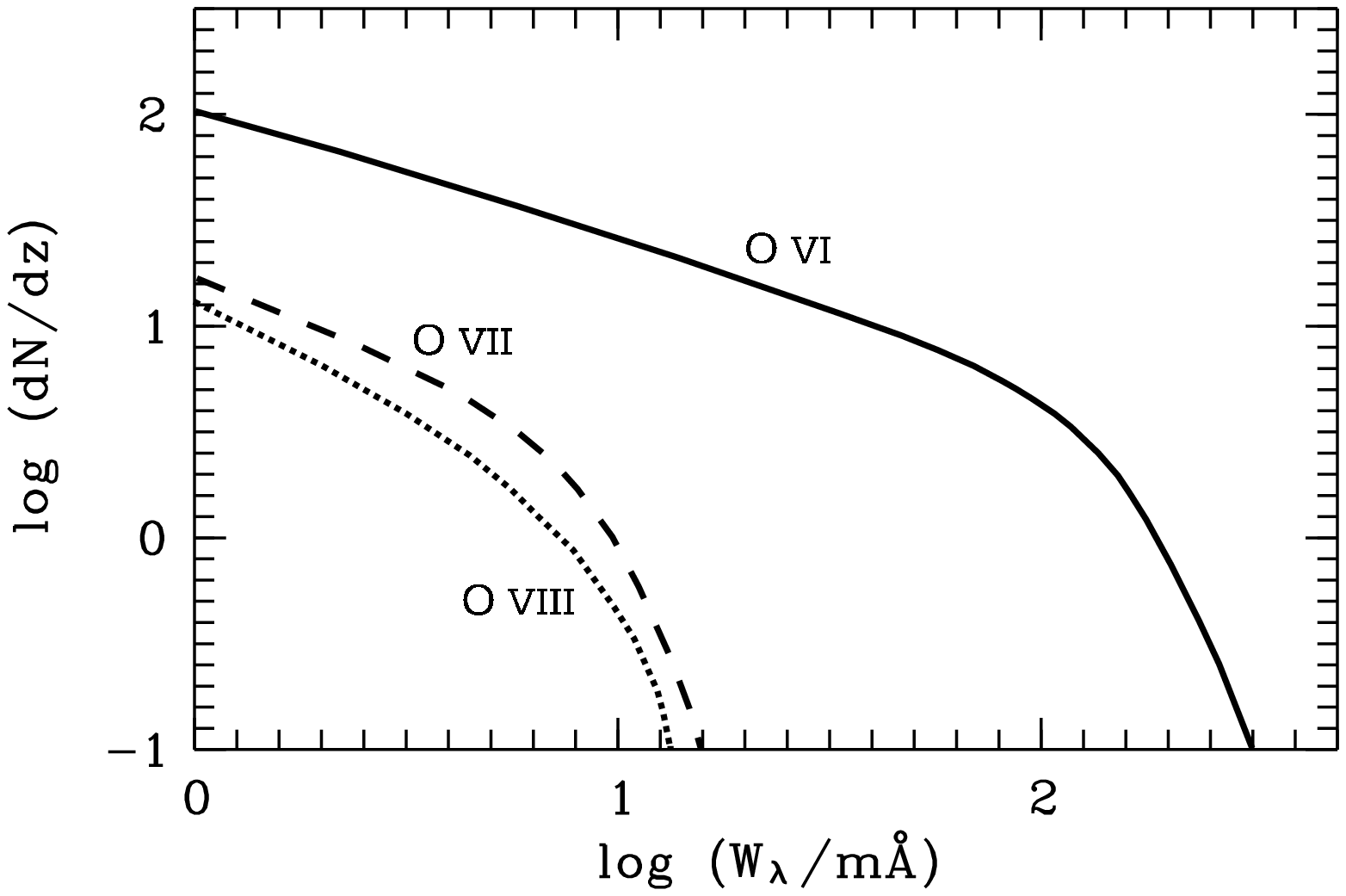}
\caption{The differential number of intervening oxygen high-ion 
(\ion{O}{vi}, \ion{O}{vii},\ion{O}{viii}) absorbers in the WHIM in a cosmological simulation is 
plotted against the equivalent width of the absorption 
(for details see \protect\citealt{cen2006}). 
While for \ion{O}{vii} and \ion{O}{viii} no significant observational results are available 
to be compared with the simulated spectra (see Sect.\,4.2),
the predicted frequency of \ion{O}{vi} absorbers is in good agreement with the observations (Sect.\,3.2). Adapted from \protect\citet{cen2006}.
}
\label{fig:fig8}
\end{center}
\end{figure}

Cosmological simulations not only have been used to investigate the large-scale distribution and physical state of the warm-hot intergalactic medium, they also have
been applied to predict statistical properties of high-ion absorption systems that can be readily compared with the UV and X-ray measurements (e.g., 
\citealt{cen2001}; \citealt{fang2001a}; \citealt{chen2003}; \citealt{furlanetto2005}; \citealt{tumlinson2005}; \citealt{cen2006}). 
Usually, a large number of artificial spectra along random sight-lines through the simulated volume are generated. Sometimes, instrumental
properties of existing spectrographs and noise characteristics are modelled, too (e.g., \citealt{fangano2007}). The most important quantities derived from
artificial spectra that can be compared with observational data are the cumulative and differential number densities (${\mathrm d}N/{\mathrm d}z$) of \ion{O}{vi},
\ion{O}{vii}, \ion{O}{viii} systems as a function of the absorption equivalent width. An example for this is shown in Fig.~\ref{fig:fig8}. Generally, there is a good
match between the simulations and observations for the overall shape of the ${\mathrm d}N/{\mathrm d}z$ distribution (see also Sect.~3.2), but mild discrepancies exist
at either low or high equivalent widths, depending on what simulation is used (see, e.g., \citealt{tripp2007}). For the interpretation of such discrepancies
it is important to keep in mind that the different simulations are based on different physical {\it models} for the gas, e.g., some simulations include galaxy feedback
models, galactic wind models, non-equilibrium ionisation conditions, etc., others do not. For more information on numerical simulations of the WHIM see 
\citealt{bertone2008} - Chapter 14, this volume.

\begin{figure}    
\begin{center}
\includegraphics[width=\hsize]{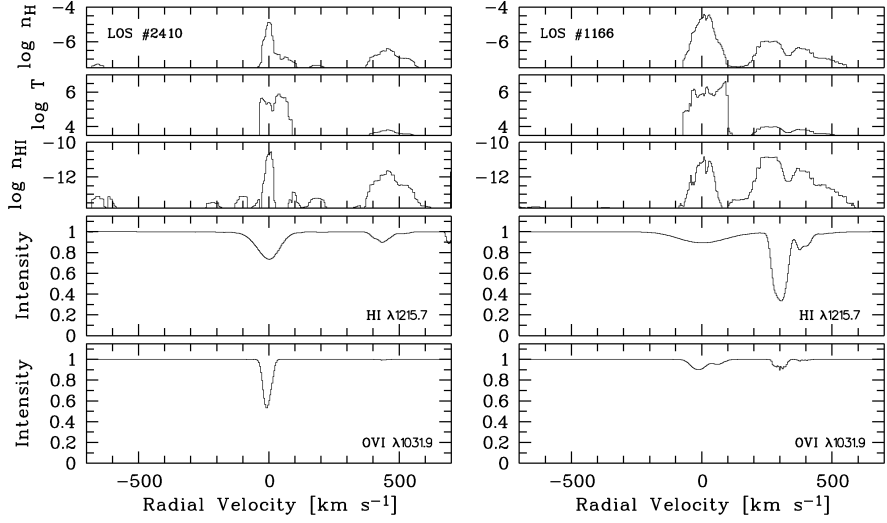}
\caption{Two examples for BLA absorbers from the WHIM in a cosmological simulation are shown. 
The panels show the logarithmic total hydrogen volume density, gas temperature, 
neutral hydrogen volume density, and normalised intensity for \ion{H}{i} Ly$\alpha$ and \ion{O}{vi} $\lambda 1031.9$ 
absorption as a function of the radial velocity along each sightline. From \protect\citet{richter2006b}.}
\label{fig:fig9}
\end{center}
\end{figure}

WHIM simulations also have been used to investigate the frequency and nature of BLAs at low redshift \citep{richter2006b}). As the simulations suggest, BLAs indeed host
a substantial fraction of the baryons at $z=0$. From the artificial UV spectra generated from their simulation Richter et al.\,derive a number of BLAs per unit redshift
of $({\mathrm d}{N}/{\mathrm d}z)_{\rm BLA}\approx 38$ for \ion{H}{i} absorbers with log $(N$(cm$^{-2})/b$(km\,s$^{-1}))\geq 10.7$,  $b\geq40$ km\,s$^{-1}$, and total
hydrogen column densities $N$(\ion{H}{ii}$)\leq 10^{20.5}$ cm$^{-2}$. The baryon content of these systems is $\sim 25$ percent of the total baryon budget in the
simulation. These results are roughly in line with the observations if partial photoionisation of BLAs is taken into account (\citealt{richter2006a};
\citealt{lehner2007}). From the simulation further follows that BLAs predominantly trace shock-heated collisionally ionised WHIM gas at temperatures log
$T\approx 4.4-6.2$. Yet, about 30 percent of the BLAs in the simulation originate in the photoionised Ly$\alpha$ forest (log $T<4.3$) and their large line widths are
determined by non-thermal broadening effects such as unresolved velocity structure and macroscopic turbulence. Fig.~\ref{fig:fig9} shows two examples of the velocity
profiles of BLAs generated from simulations presented in \citet{richter2006b}.

The results from the analysis of artificially generated UV spectra underline that the comparison between WHIM simulations and quasar absorption line studies indeed are
quite important for improving both the physical models in cosmological simulations and the strategies for future observations of the warm-hot intergalactic gas.

\subsection{WHIM absorbers at high redshift}\label{WHIM absorbers at high redshift}

Although this chapter concentrates on the properties of WHIM absorbers at low redshift (as visible in UV and X-ray absorption) a few words about high-ion absorption at
high redshifts ($z>2$) shall be given at this point. At redshifts $z>2$, by far most of the baryons are residing in the photoionised intergalactic medium that gives
rise to the Ly$\alpha$ forest. At this early epoch of the Universe, baryons situated in galaxies and in warm-hot intergalactic gas created by large-scale structure
formation contribute together with only $<15$ percent to the total baryon content of the Universe. Despite the relative unimportant role of the WHIM at high $z$,
\ion{O}{vi} absorbers are commonly found in optical spectra of high-redshift quasars (e.g., \citealt{bergeron2002}; Carswell et al. \citealt{carswell2002};
Simcoe et al. \citealt{simcoe2004}). The observation of intervening \ion{O}{vi} absorbers at high redshift is much easier than in the local Universe, since the absorption
features are redshifted into the optical regime and thus are easily accessible with ground-based observatories. However, blending problems with the numerous \ion{H}{i}
Ly$\alpha$ forest lines at high $z$ are much more severe than for low-redshift sightlines. Because of the higher intensity of the metagalactic UV background at high
redshift it is expected that many of the \ion{O}{vi} systems in the early Universe are photoionised. Collisional ionisation of \ion{O}{vi} yet may be important for
high-redshift absorbers that originate in galactic winds (see, e.g., \citealt{fangano2007}). While for low redshifts the population of \ion{O}{vi}
absorbers is important for the search of the "mission baryons" that are locked in the WHIM phase in the local Universe, \ion{O}{vi} absorbers at high redshift are
believed to represent a solution for the problem of the "missing metals" in the early epochs of structure formation. This problem arises from the facts that at high
redshift an IGM metallicity of $\sim 0.04$ is expected from the star-formation activity of Lyman-Break Galaxies (LBGs), while observations of intervening \ion{C}{iv}
absorption systems suggest an IGM abundance of only $\sim 0.001$ solar (\citealt{songaila2001}; \citealt{scannapieco2006}), thus more than one
order of magnitude too low. Possibly, most of the missing metals at high $z$ are hidden in highly-ionised hot gaseous halos that surround the star-forming galaxies
(e.g., \citealt{ferrara2005}) and thus should be detectable only with high ions such as \ion{O}{vi} rather than with intermediate ions such as \ion{C}{iv}.
Using the UVES spectrograph installed on the {\it Very Large Telescope} (VLT) \citet{bergeron2005} have studied the properties of high-redshift
\ion{O}{vi} absorbers along ten QSO sightlines and have found possible evidence for such a scenario. Additional studies are required to investigate the nature of
high-$z$ \ion{O}{vi} systems and their relation to galactic structures in more detail. However, from the existing measurements clearly follows that the study of
high-ion absorbers at large redshifts is of great importance to our understanding of the formation and evolution of galactic structures at high $z$ and the transport of
metals into the IGM.

\subsection{Concluding remarks}\label{Concluding remarks}

The analysis of absorption features from high ions of heavy elements and neutral hydrogen currently represents the best method to study baryon content, physical properties, and distribution of the warm-hot intergalactic gas in large-scale filaments at low and high redshift. However, the interpretation of these spectral signatures in terms of WHIM baryon content and origin still is afflicted with rather large systematic uncertainties due to the limited data quality and the often poorly known physical conditions in WHIM absorbers (e.g., ionisation conditions, metal content, etc.). Future instruments in the UV (e.g., COS) and in the X-ray band (e.g., {\it XEUS}, {\it Constellation X}) hold the prospect of providing large amounts of new data on the WHIM with good signal-to-noise ratios and substantially improved absorber statistics. These missions therefore will be of great importance to improve our understanding of this important intergalactic gas phase.

\begin{acknowledgements}
The authors thank ISSI (Bern) for support of the team "Non-virialized X-ray components in clusters of galaxies". SRON is supported fiancially by NWO, the Netherlands Organization for Scientific Research.
\end{acknowledgements}

 \end{document}